\documentclass[conference]{IEEEtran}
\IEEEoverridecommandlockouts
\usepackage{cite}
\usepackage{amsmath,amssymb,amsfonts}
\usepackage{algorithmic}
\usepackage{graphicx}
\usepackage{textcomp}
\usepackage{xcolor}
\def\BibTeX{{\rm B\kern-.05em{\sc i\kern-.025em b}\kern-.08em
    T\kern-.1667em\lower.7ex\hbox{E}\kern-.125emX}}
    
    \usepackage{amsthm}
    
    \newtheorem{theorem}{Theorem}[section]

\newtheorem{remark}[theorem]{Remark}

\usepackage{tikz}
\usepackage{amsmath}
\usepackage{color}
\usepackage{balance}
\usepackage{afterpage}
\usepackage{xspace}
\usepackage[font=small]{caption} 
\usepackage[labelformat=simple]{subcaption}

\usepackage{algorithm}
\usepackage{algorithmic}
\usepackage{graphicx}
\usepackage{multirow}
\usepackage{breakurl}
\PassOptionsToPackage{hyphens}{url}\usepackage{hyperref}

\hypersetup{
    colorlinks=true,
    linkcolor=magenta,
    filecolor=magenta,      
    urlcolor=magenta,
    citecolor=magenta,
}

\usepackage{amsthm,amsfonts}

\usepackage{enumitem}
\usepackage{booktabs}

\ifCLASSOPTIONcompsoc
\usepackage[nocompress]{cite}
\else
\usepackage{cite}
\fi

\definecolor{forestGreen}{RGB}{34,139,34}
\newcommand{\tian}[1]{{\color{red}\textbf{Tian:} \textit{#1}}}

\newcommand{\shijian}[1]{{\color{cyan}\textbf{shijian:}
\textit{#1}}}
\newcommand{\Lijie}[1]{{\color{red}\textbf{Lijie:} \textit{#1}}}

\newcommand{\1}{{\em (i)}}
\newcommand{\2}{{\em (ii)}}
\newcommand{\3}{{\em (iii)}}
\newcommand{\4}{{\em (iv)}}
\newcommand{\5}{{\em (v)}}

\newcommand{\para }[1]{\smallskip \noindent  {\bf {#1}}}
\newcommand{\sysname}{\emph{Sync-Switch}\xspace}
\newcommand{\eat}[1]{}

\begin{document}

\bstctlcite{IEEEexample:BSTcontrol}

\title{\sysname: Hybrid Parameter Synchronization for Distributed Deep Learning
}

\author{\IEEEauthorblockN{Shijian Li, 
Oren Mangoubi, 
Lijie Xu\IEEEauthorrefmark{2} and 
Tian Guo}
\IEEEauthorblockA{Worcester Polytechnic Institute
}
\IEEEauthorblockA{\IEEEauthorrefmark{2}State Key Lab of Computer Science, Institute of Software, Chinese Academy of Sciences
}
}

\maketitle

\thispagestyle{plain}
\pagestyle{plain}

\begin{abstract}

Stochastic Gradient Descent (SGD) has become the de facto way to train deep neural networks in distributed clusters.  
A critical factor in determining the training throughput and model accuracy is the choice of the parameter synchronization protocol.
For example, while Bulk Synchronous Parallel (BSP) often achieves better converged accuracy, the corresponding training throughput can be negatively impacted by stragglers. 
In contrast, Asynchronous Parallel (ASP) can have higher throughput, but its convergence and accuracy can be impacted by stale gradients. 
To improve the performance of synchronization protocol, recent work often focuses on designing new protocols with a heavy reliance on hard-to-tune hyper-parameters. 

In this paper, we design a hybrid synchronization approach that exploits the benefits of both BSP and ASP, i.e., reducing training time while simultaneously maintaining the converged accuracy. 
Based on extensive empirical profiling, we devise a collection of adaptive policies that determine how and when to switch between synchronization protocols. 
Our policies include both offline ones that target recurring jobs and online ones for handling transient stragglers.
We implement the proposed policies in a prototype system, called \sysname, on top of TensorFlow, and evaluate the training performance with popular deep learning models and datasets.
Our experiments show that \sysname achieves up to 5.13X throughput speedup and similar converged accuracy when comparing to BSP. Further, we observe that \sysname achieves 3.8\% higher converged accuracy with just 1.23X the training time compared to training with ASP. Moreover, \sysname can be used in settings when training with ASP leads to divergence errors. 
\sysname achieves all of these benefits with very low overhead, e.g., the framework overhead can be as low as 1.7\% of the total training time. 

\end{abstract}


\section{Introduction}
\label{sec:intro}


We are witnessing the increasingly widespread adoption of deep learning in a plethora of application domains. 
The unprecedented success of deep learning is, in large part, powered by rapid model innovations, which in turn critically depend on algorithms and systems support for training.
One of these innovations, \emph{distributed deep learning}--training deep neural networks on a cluster of GPU servers--is increasingly leveraged to train complex models on larger datasets.
In particular, SGD-based optimization 
has emerged as the de facto way to perform distributed training and provides the basis for parallelizing training jobs, allowing deep learning practitioners to evaluate different model variants quickly.

However, it is more difficult to achieve good performance and high-quality training with SGD-based distributed training, compared to traditional single-node training~\cite{shi2018performance, Li2019-oe}. A large number of factors, 
such as slow servers and network communication links,  
 can all impact the distributed training performance~\cite{Yan2015-jv,Ben-Nun2019-ca,Li2020-bk}. Of particular importance is how each cluster node communicates and synchronizes their respective progress during training, governed by \emph{parameter synchronization protocols}, which has a profound impact on both the model converged accuracy and training time. 
Bulk synchronous parallel (BSP)~\cite{Gerbessiotis1994-wy}, a default option for popular frameworks including TensorFlow, requires each node to synchronize every iteration. In contrast, asynchronous parallel (ASP) allows nodes to work at their own pace~\cite{dean2012large}. 
However, both distributed training protocols have their respective limitations, e.g., BSP is prone to slow down due to workers need to wait for synchronization
while ASP suffers from decreased accuracy due to stale gradients.

\begin{figure}[t]
    \centering
    \includegraphics[width=0.35\textwidth]{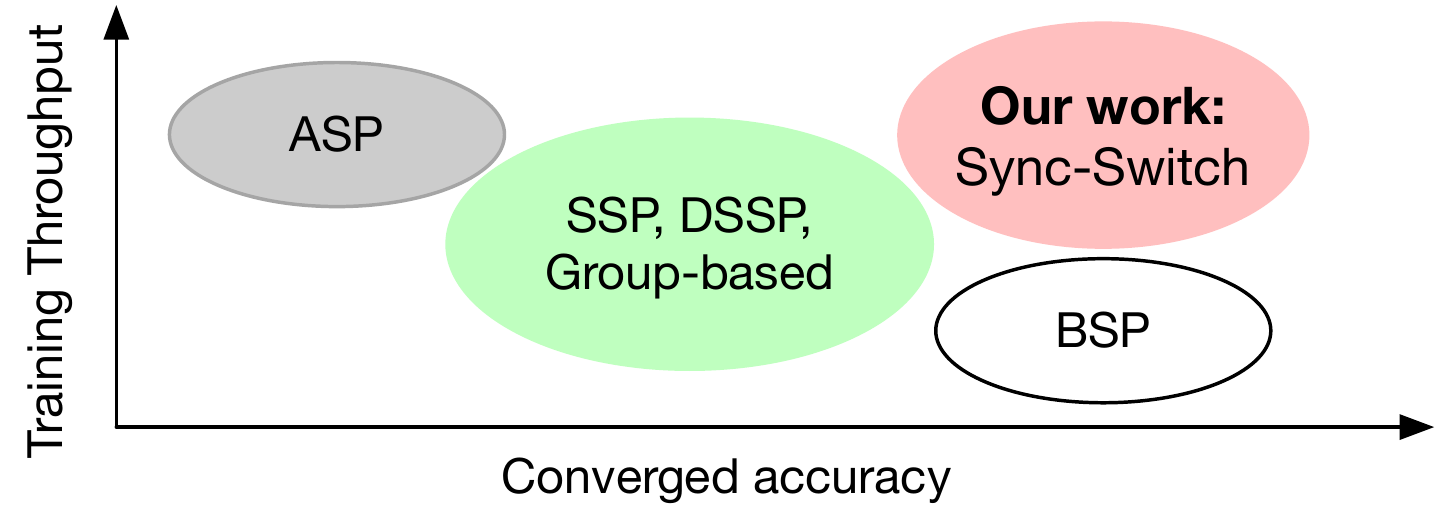}
    \caption{
    \textbf{Ours vs. prior work on synchronization protocols.}
    Our work looks to improve both the training time and accuracy simultaneously, compared to prior work that trades-off these two metrics.
    }
    \label{fig:sync-compare}
    \vspace{-1em}
\end{figure}


In this work, we explore ways to exploit the benefits of both BSP and ASP and design a hybrid synchronization approach \sysname. 
In contrast to prior work~\cite{zhao2019dynamic,jiang2019novel,hsieh2017gaia,dutta2020slow}, which often needs to sacrifice either training throughput or converged accuracy, we set out to reduce training time while simultaneously maintaining the converged accuracy
as illustrated in \figurename~\ref{fig:sync-compare}. 
Specifically, we propose an empirically-driven methodology---which we also use to generate a set of policies---that determine how and when to switch the synchronization protocol. 
%
%
Our policies, the offline ones that target jobs under normal training circumstances and the online ones that react to cluster runtime status, are designed with the key insight of maximizing the time the GPU servers spend on training asynchronously without sacrificing the trained model's accuracy. 


\begin{figure}[t!]
    \centering
    \begin{subfigure}[t]{0.24\textwidth}
        \includegraphics[width=\textwidth]{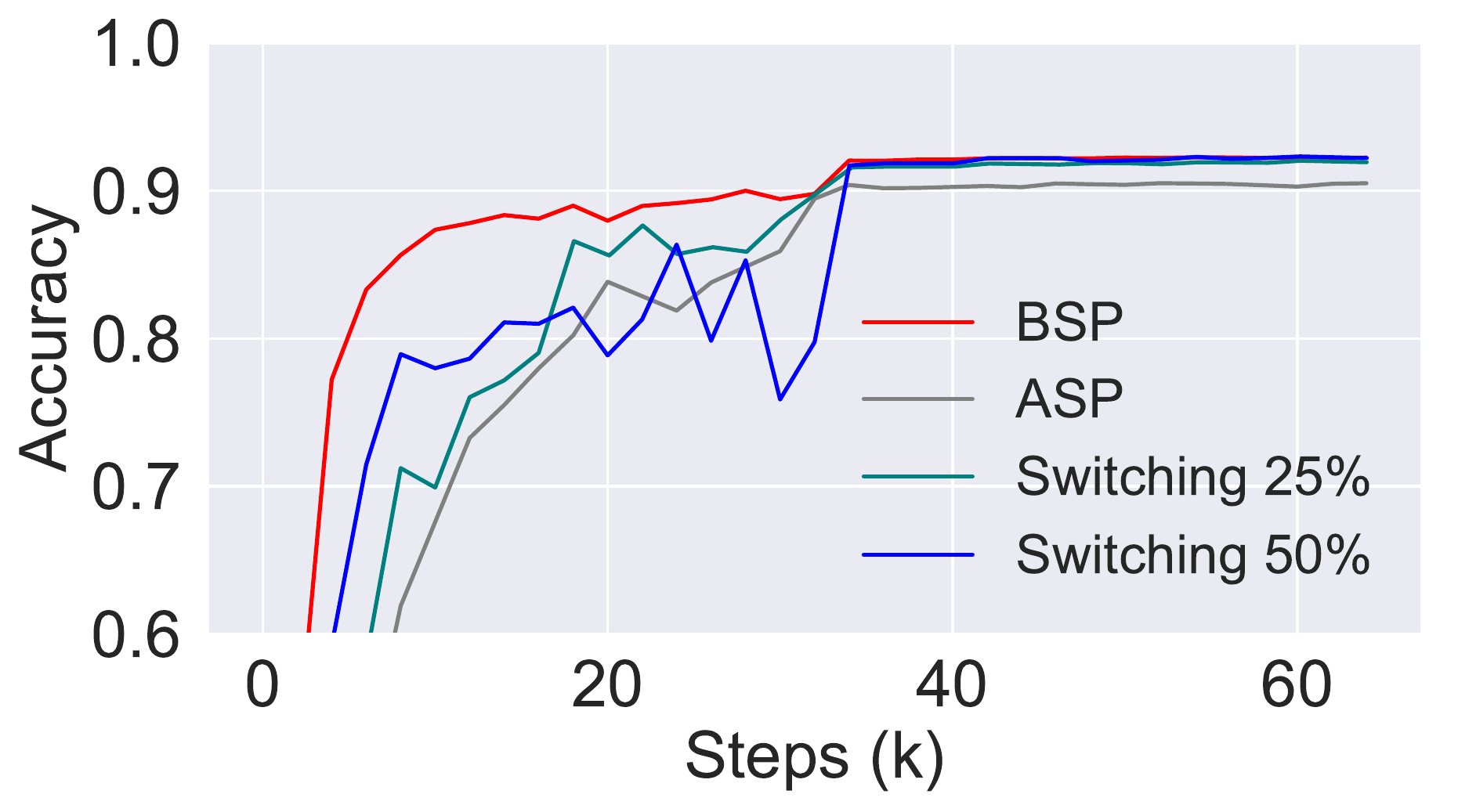}
        \caption{Test accuracy.}
        \label{fig:sync_teaser_results:accuarcy}
    \end{subfigure}
    \hfill
    \begin{subfigure}[t]{0.24\textwidth}
        \includegraphics[width=\textwidth]{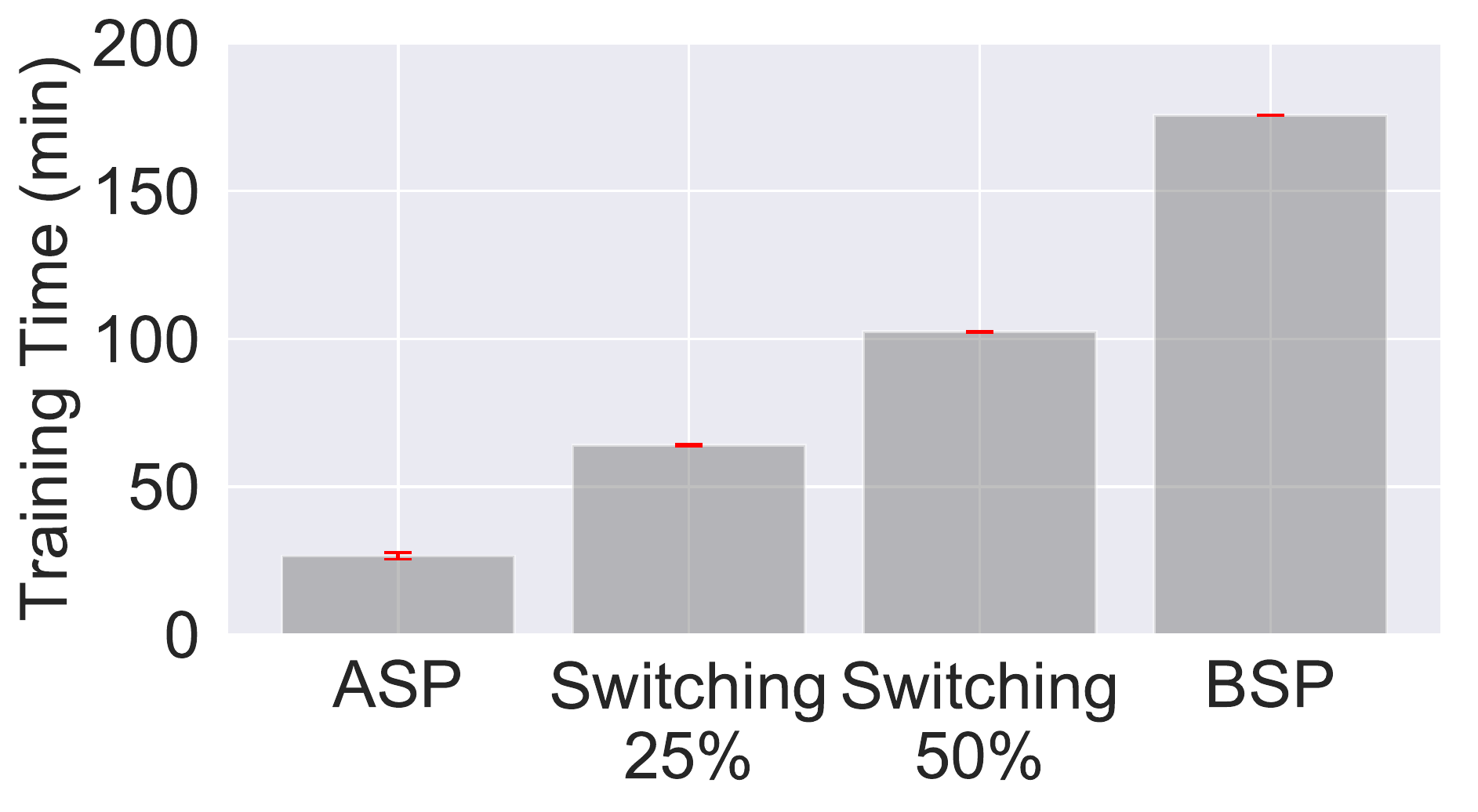}
        \caption{Total training time.}
        \label{fig:sync_teaser_results:time}
    \end{subfigure}
    \caption{
    \textbf{Benefits of synchronization switching.}
    Training ResNet32 on Cifar-10 using 8 machines, with BSP then ASP,  reduces the training time by up to 63.5\% while achieving similar converged accuracy, compared to training with BSP.
    }
    \label{fig:sync_teaser_results}
    \vspace{-1em}
\end{figure}


Through extensive empirical profiling of distributed training workload, we demonstrate that our key idea of \emph{hybrid synchronization} can lead to converged accuracy and training time benefit compared to using BSP and using ASP, respectively.  
In particular, we empirically observe that, while one may need to perform synchronous training throughout all epochs to achieve the highest possible {\em training} accuracy, only a minimal amount of synchronous training is needed to achieve a high-quality {\em test} accuracy. Indeed, as shown in Figure~\ref{fig:sync_teaser_results:accuarcy}, starting the training synchronously with BSP and then switching to an asynchronous protocol achieves almost the same converged test accuracy compared to training exclusively with BSP. Furthermore, by spending less time training with BSP (50\% of training workload vs. 25\%), the total training time is reduced by 37.5\% (see Section~\ref{subsec:eval_methodology} for detailed methodology). 

To determine \emph{how and when to use} BSP and ASP synchronizations, we  develop an empirical-driven methodology that allows us to derive policies for a given distributed training workload.
Specifically, by comparing the training performance under different settings, one can derive the first \emph{protocol policy} that describes the relative execution order of the synchronization protocols. 
We found that training with BSP (a more precise computation method) first and then switching to ASP (a less precise one) leads to improved training throughput and similar converged accuracy, compared to training with BSP.
To determine when to use BSP and when to use ASP, we introduce both online and offline \emph{timing policies}. 
Based on our empirical observation that training longer with BSP does not improve converged accuracy beyond a knee point, we use a binary search-based approach to find the optimal switch timing.
Furthermore, to account for transient stragglers, we devise an online policy that works in tandem with the optimal timing policy obtained offline.
Finally, to properly adjust hyper-parameters when switching to a new synchronization protocol, we devise the \emph{configuration policies} by adapting prior work and by taking into account factors including the cluster size and the training stage.

We implement the proposed policies in a prototype system called \sysname on top of a popular distributed training framework TensorFlow and evaluate the training performance under three distributed training setups.
Our experiments show that \sysname achieves up to 5.13X throughput speedup and similar converged accuracy when comparing to BSP. Further, we observe that \sysname achieves 3.8\% higher converged accuracy with just 1.23X the training time compared to training with ASP. Moreover, \sysname can be used in settings when training with ASP leads to divergence errors. 
\sysname is also effective in handling transient stragglers (nodes that exhibit temporary slowdown), mitigating the potential performance degradation. 
Furthermore, we quantify the overhead of using \sysname in terms of offline search cost and runtime overhead. Specifically, we show that the upfront search cost can be quickly amortized for jobs with more than ten recurrences, a very likely scenario given the trial-and-error nature of deep learning training. Additionally, \sysname incurs as low as 36 seconds, or about 1.7\% of the total training time, overhead in switching between BSP and ASP protocols.

In summary, we make the following main contributions: 
\begin{itemize}[leftmargin=.12in,topsep=0pt]
\item 
We propose a methodology and derive policies that govern hybrid parameter synchronization to improve the training throughput while \emph{simultaneously} maintaining high-quality converged accuracy. 
The offline policies can be derived for a given distributed training workload and are particularly useful for recurring jobs\footnote{Recurring jobs refer to training jobs with the same deep learning model, but can be trained with different hyper-parameters, such as hyper-parameter tuning, or different datasets, such as online learning.},
while the online policies are effective in dealing with stragglers that are temporary. 
%
\item We implement a prototype system called \sysname, on top of a popular deep learning framework TensorFlow, that encapsulates all the adaptive policies. Deep learning practitioners can directly leverage \sysname without modifying the distributed training scripts. Our code and experiment data are available in the project GitHub repository: \url{https://github.com/cake-lab/Sync-Switch}.
\item 
We demonstrate the efficacy of these policies through experiments using popular convolutional neural networks and datasets for image classification. We show that \sysname can improve the total training time by up to 5X while achieving similar test accuracy, compared to training with BSP.
\sysname also achieves 4X improvement on the time-to-accuracy metric, outperforming prior work that reported a speedup of 1.1X-2X when using protocols SSP and DSSP~\cite{zhao2019dynamic}.
Further, \sysname can effectively circumvent the performance degradation caused by transient stragglers under moderate slowdown scenarios.
\end{itemize}


\section{Background}
\label{sec:background}

\begin{figure*}[t]
    \centering
    \includegraphics[width=0.95\textwidth]{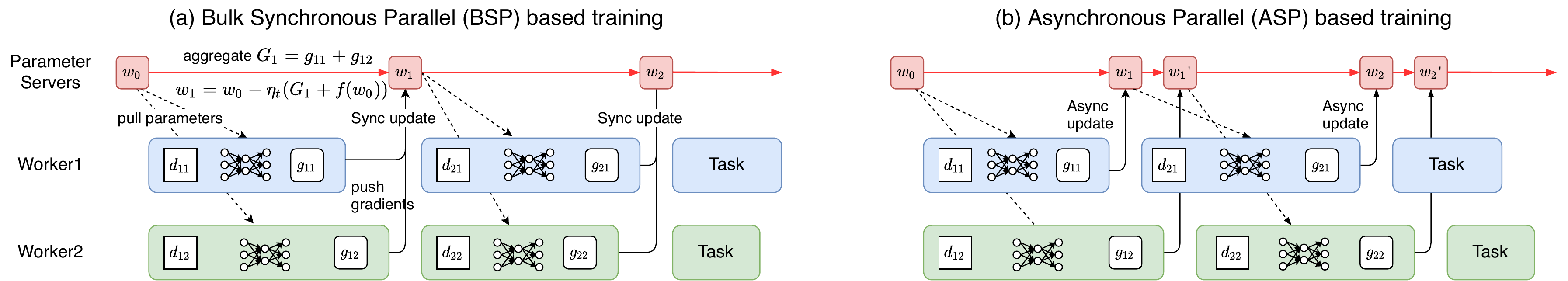}
    \caption{
    \textbf{Popular distributed parameter synchronization protocols.} For BSP, parameter servers use barriers for gradient collection and only start to update parameters $w$ once all gradients $g$ are aggregated. In contrast, ASP workers push/pull gradients/parameters at their own pace.
    }
    \label{fig:distributed_synchronization_protocols}
    \vspace{-1em}
\end{figure*}


\subsection{Distributed Deep Learning}
\label{subsec:dll}

In this work, we target \emph{distributed deep learning} where multiple GPU-equipped computing nodes work together to train a deep learning model by communicating gradients and parameters over network. We focus on the more popular approach \emph{data parallelism}, where the training data are partitioned and offloaded to the workers, instead of model parallelism where models themselves are distributed.
When training with data parallelism, each worker trains on the complete neural network in \emph{steps}, which each step corresponds to going through one mini-batch of data,
to update the model parameters. 

Further, we focus on \emph{parameter server (PS)} based architecture that consists of two logical entities: PS and worker. 
We choose to collocate PSs and workers, to better exploit computational resources and to reduce network communication, and configure the training cluster with equal numbers of PSs and workers based on prior work~\cite{Peng2018-mw}. 
To train a neural network, a worker will first pull model parameters from all PSs, then perform the forward and backward propagation computation on a mini-batch and the current model parameters, and finally push the computed gradients to all PSs. Depending on the parameter synchronization protocol in use, the PSs will immediately update the model parameters or wait until receiving all gradients from all workers. 
%

\subsection{Distributed Parameter Synchronization Protocols}
\label{subsec:protocols}

\begin{figure}[t]
    \centering
    \begin{subfigure}[t]{0.24\textwidth}
        \includegraphics[width=\textwidth]{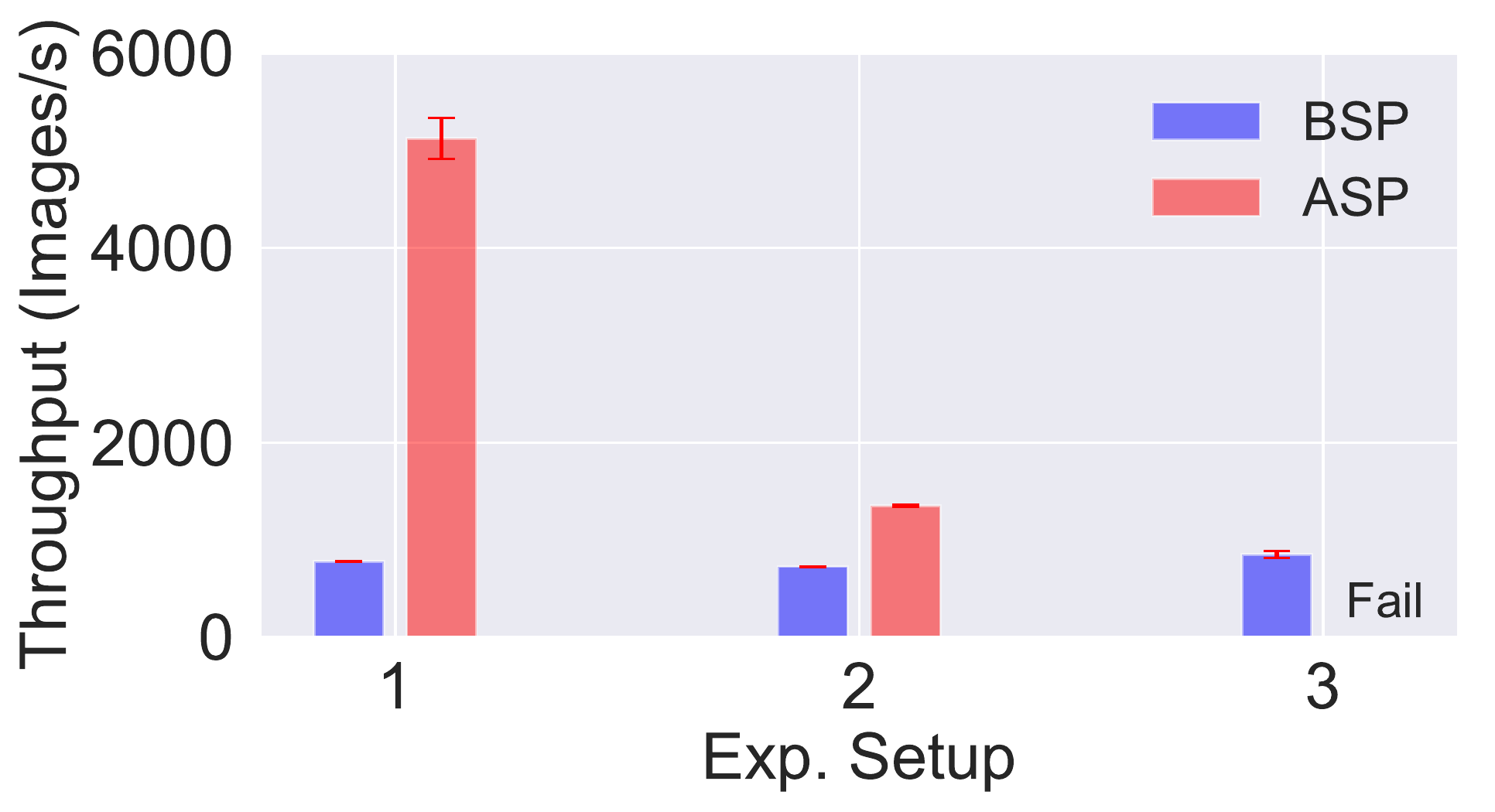} 
        \caption{Without straggler. 
        }
        \label{fig:bsp_asp_throughput}
    \end{subfigure}
    \hfill
    \begin{subfigure}[t]{0.24\textwidth}
        \includegraphics[width=\textwidth]{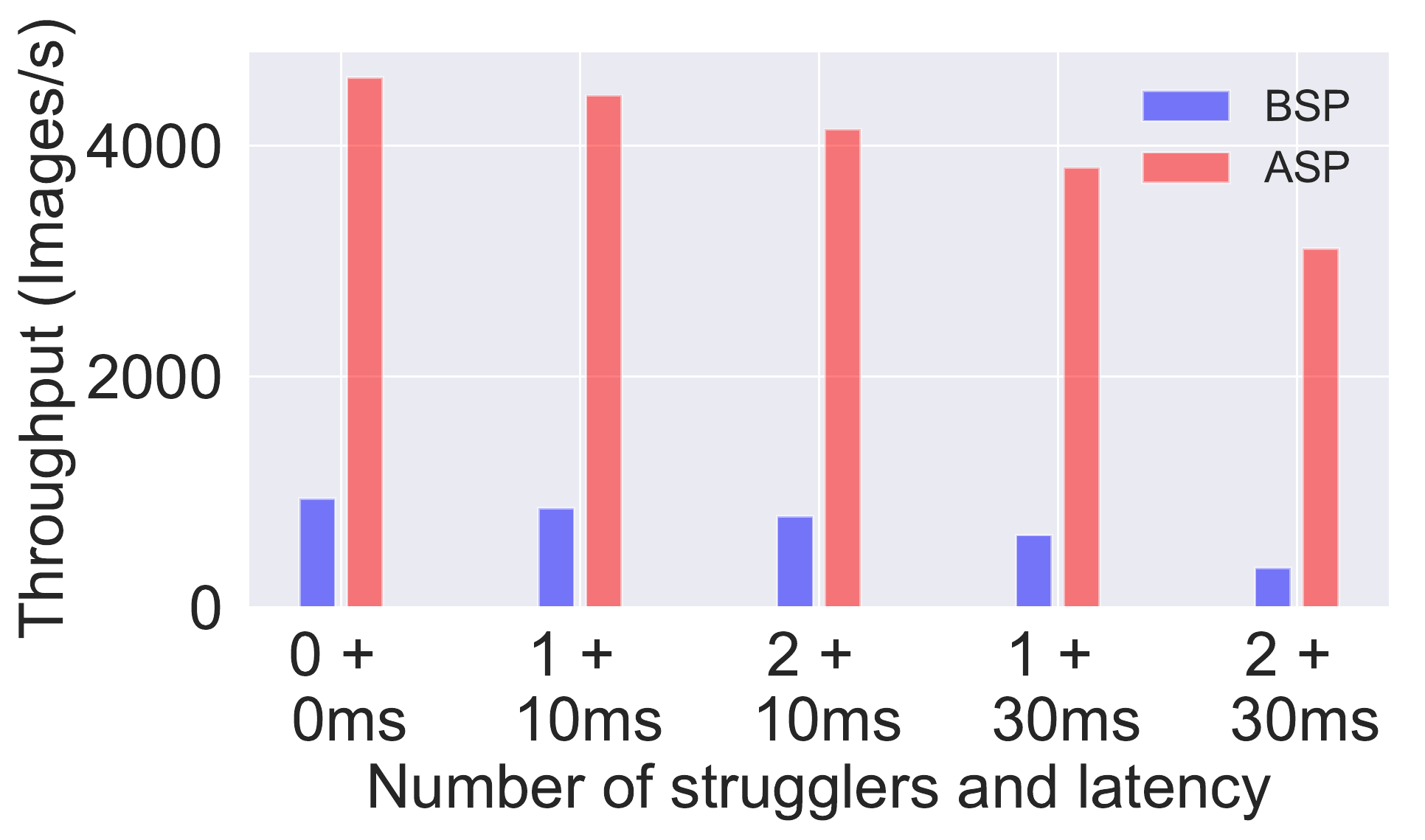} 
        \caption{Exp. Setup 1: with straggler.}
        \label{fig:bsp_asp_throughput_straggler}
    \end{subfigure}
    \caption{\textbf{Training throughput comparison between BSP and ASP.} We observe that training with ASP has higher throughput than with BSP, even without stragglers. See Section~\ref{subsec:eval_methodology} for a detailed methodology.
    }
    \label{fig:exp_bsp_vs_asp}
    \vspace{-1em}
\end{figure}

In distributed deep learning, synchronization and coordination of worker progress (i.e., gradient computation) is achieved by \emph{distributed parameter synchronization protocols}~\cite{Gerbessiotis1994-wy,Recht2011-bh,dean2012large,zhao2019dynamic,dutta2020slow}. There are two popular protocols:
\emph{Bulk Synchronous Parallel} (BSP) and \emph{Asynchronous Parallel} (ASP) that differ in their horizontal scalability,
sensitivity to stragglers, and converged test accuracy. 
%
Via an empirical measurement with different training workloads and cluster sizes,   
we quantify the training throughput difference between BSP and ASP as shown in Figure~\ref{fig:exp_bsp_vs_asp}. 
Given the same experiment setup, training with ASP can be up to 6.59X faster than with BSP, especially when stragglers are presented. Our observations align with recent work~\cite{Or2020-nh,Peng2018-mw} and suggest the promise of leveraging ASP to improve training time. 
%


BSP, as shown in Figure~\ref{fig:distributed_synchronization_protocols}(a), is a deterministic scheme where workers perform a parallel computation phase followed by a synchronization phase where the gradients (e.g., $g_{11}$ and $g_{12}$) are aggregated at a barrier. The model parameters are then updated according to this aggregated set of gradients. This method ensures that all workers work on the same model parameters and prevents any workers from proceeding to the next mini-batch.
%
This synchronous update guarantees the method to be equivalent to a true mini-batch stochastic gradient descent algorithm, 
which can ensure the correctness of the parallel computation~\cite{chen2016revisiting}.
%
Since BSP workers need to wait for all updates to the model parameters
at the barrier, the faster workers all remain idle while waiting for the
stragglers. This drastically limits the training performance of the whole
training cluster. Generally speaking, BSP offers high converged accuracy but suffers from computation inefficiency, especially in unfavorable  environments. 

ASP, as shown in Figure~\ref{fig:distributed_synchronization_protocols}(b), allows computations to execute in parallel as fast as possible by running workers completely asynchronously. Each worker individually pulls parameters and pushes new gradients for updates at the parameter server. For example, once Worker1 pushes the computed gradient $g_{11}$ to the PSs, $w_0$ is updated as $w_1 = w_0 - \eta_t(g_{11} + f(w_0))$ and used by the subsequent task in Worker1. Similarly, the PSs will update $w_1$ as $w_1' = w_1 - \eta_t(g_{12} + f(w_1))$ when receiving $g_{12}$ from Worker2. 
Consequently, ASP can often achieve better speedups than BSP.
However, ASP can be impacted by the stale gradients: 
the tasks may use old model parameters (e.g., the second tasks of Worker1 and Worker2 use different model parameters $w_1$ and $w_1'$) for training, which introduces noise and error into the computation.
As such, the model trained with ASP often converges to lower training and test accuracy when compared to BSP~\cite{chen2016revisiting, dutta2020slow}. 


\section{Problem Statement and Solution Overview}
\label{sec:motivation}

In this paper, we investigate \emph{how to improve the training throughput while simultaneously maintaining the converged accuracy} for distributed deep learning. 
Our study is motivated by the inherent limitations of existing synchronization protocols, as previously discussed in Section~\ref{sec:background}. Our key insight is that by adaptively switching between the synchronization protocols based on both internal training status and runtime factors, we can avoid their respective drawbacks such as low speedup and low converged accuracy as much as possible. 

\para{System model.}
We consider the scenario of training a deep learning model in a dedicated cloud-based GPU cluster. We focus on the popular parameter-server-based distributed training architecture adopted by TensorFlow and prior work~\cite{Peng2018-mw,Li2020-bk}, and collocate the PS and worker on the same physical server. Further, we target the training of deep convolutional neural networks on widely used image datasets with data parallelism, a commonly used approach for models that can fit into the memory of a discrete GPU. We assume deep learning practitioners will provide a training script that specifies the initial training configuration, describing the training cluster and the deep learning workload, as well as the hyper-parameters. This assumption is reasonable as distributed training is often an iterative and recurring process, making such training scripts readily accessible.





\para{Challenges.}
There are three key challenges in designing policies for hybrid synchronization. First, given that the training itself is a stochastic process, it is challenging to leverage internal training metrics such as training loss and anytime accuracy to extrapolate and generalize observed performance benefits. 
Second, we have observed high accuracy variations, even using the same training workload and cluster setup (e.g., Figure~\ref{fig:bsp_order}). As such, it necessitates the consideration of this inherent performance fluctuation when designing any policies. Additionally, it also makes designing empirical-based policies costly at the very least because each configuration needs to be evaluated multiple times. 
Third, distributed training can be prone to runtime performance variations such as network bandwidth fluctuations, and if left undealt with, the stragglers can lead to degraded training performance (e.g., Figure~\ref{fig:exp_bsp_vs_asp}).


\para{\sysname Overview.}
We address the above-mentioned challenges of hybrid synchronization with an guided empirical exploration and introduce a new prototype system called \sysname. 
Specifically, we devise a set of policies that regulate the protocol, the timing, and the configuration to use for distributed training. Our policies include the offline ones that are generated by a binary search-based algorithm and the online ones for handling transient stragglers with temporary slowdown\footnote{This differs from a longer-term slowdown, which can be most effectively handled by replacing the slow worker~\cite{Or2020-nh}.}. These policies can be used in tandem to improve the training throughput while \emph{simultaneously} maintaining high-quality converged accuracy for distributed training jobs, as we will demonstrate empirically in Section~\ref{sec:eval}.
To support adaptive synchronization, \sysname is built upon the existing framework functionalities such as saving the training progress and restarting from the checkpoint as well as mechanisms to monitor and collect internal training metrics. More details on implementation will be described in Section~\ref{sec:impl}.

\section{\sysname Policy Design}
\label{sec:design}

In this section, we introduce a set of policies that determine \emph{how and when to switch to a different synchronization}. This includes offline policies that target recurring jobs (Sections~\ref{subsec:ordering} and \ref{subsec:offline_binary_search}), online policies that react to training status (Section~\ref{subsubsec:online_policies}), and policies for adjusting hyper-parameters.  



\subsection{Protocol Policy: Which to Use and in What Order?}
\label{subsec:ordering}


In this work, we focus on selecting between two synchronization protocols, i.e., fully synchronous (BSP) and fully asynchronous (ASP). As described previously in Section~\ref{sec:intro}, there are a plethora of other protocols that provide different trade-offs between training throughput and converged accuracy. Mixing in these in-between training protocols might produce a slightly better outcome due to larger decision spaces; however, these in-between protocols can be complicated to use due to lack of framework support and the use of extra hyper-parameters~\cite{dutta2020slow,jiang2019novel,hsieh2017gaia}.

Our first policy states that we should start the training with BSP and then switch to ASP when the following conditions are satisfied: 
\1 When switching to ASP yields a similar converged {\em test} accuracy (but perhaps a lower {\em training} accuracy) as continuing with BSP;
\2 when transient stragglers arise in the training cluster. We will describe the policies that verify both conditions to determine the switching timing in Section~\ref{subsec:timing_policy}. 

\begin{figure}[t]
    \centering
    \begin{subfigure}[b]{0.24\textwidth}
        \includegraphics[width=\textwidth]{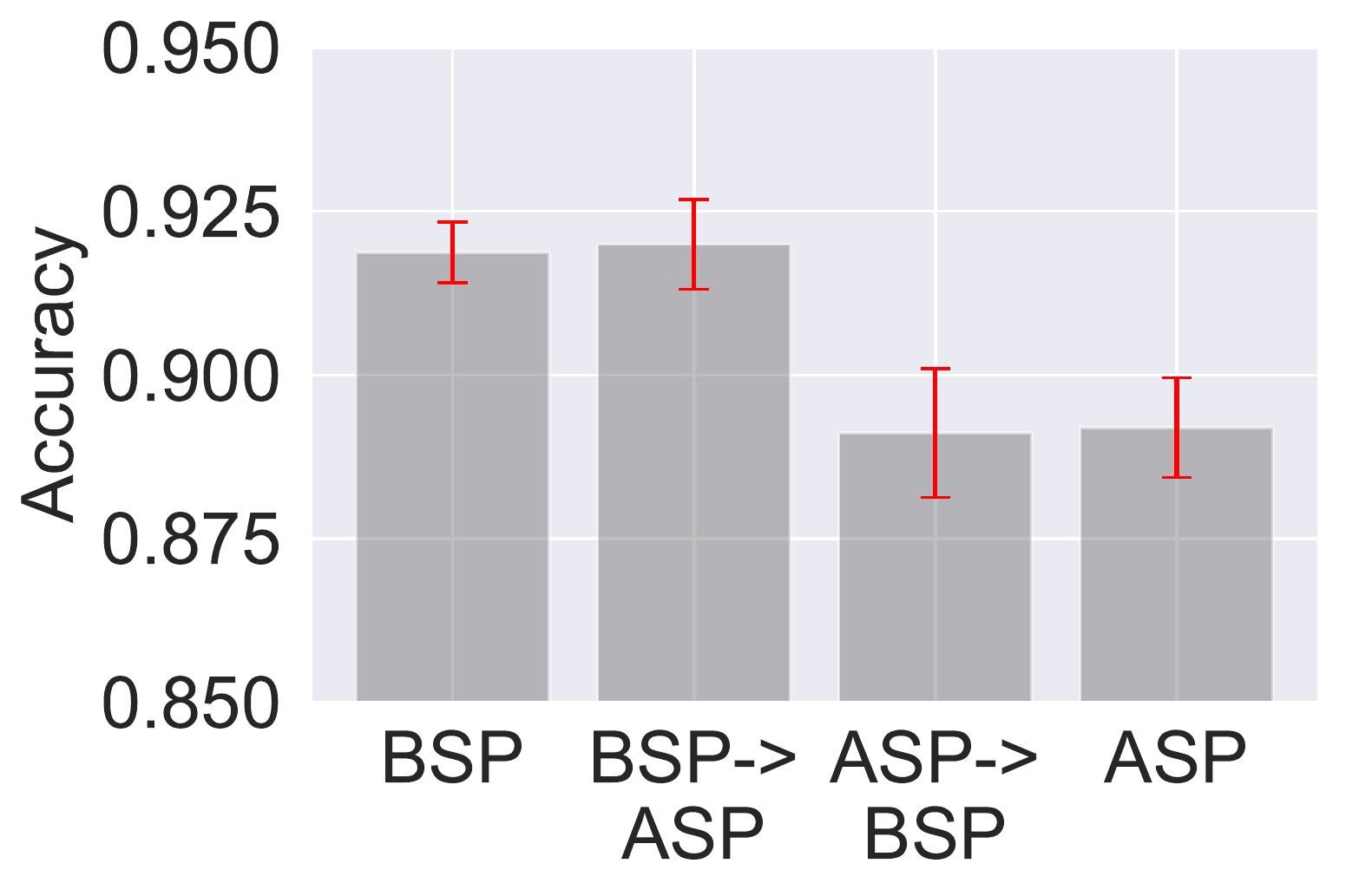} 
        \caption{Order of synchronicity.}
        \label{fig:bsp_order}
    \end{subfigure}
    \hfill
    \begin{subfigure}[b]{0.24\textwidth}
        \includegraphics[width=\textwidth]{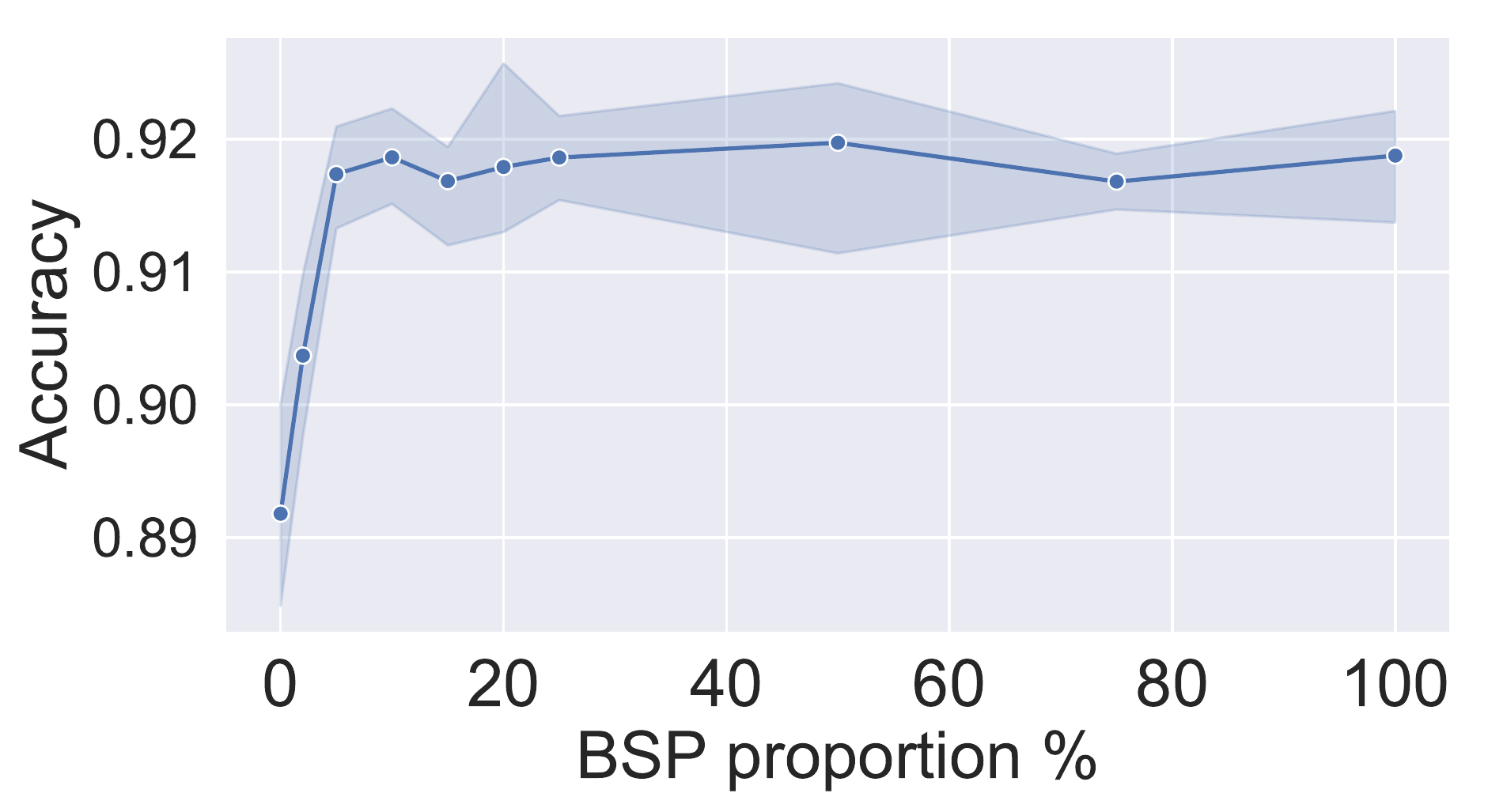} 
        \caption{Percentage of synchronicity.}
        \label{fig:bsp_percentage}
    \end{subfigure}
    \vspace{-1em}
    \caption{\textbf{Impact of synchronicity.}
    We observe that \1 switching from BSP to ASP can maintain the same level of converged accuracy and \2 training longer with BSP does not further improve accuracy beyond the knee point.
    }
    \label{fig:exp_bsp_order_percentage}
    \vspace{-1em}
\end{figure}

\subsubsection{Empirical Analysis}
To evaluate our hypothesis that running BSP at earlier epochs and switching to ASP at later epochs allows one to obtain a test error which is as low as using BSP during the entire training, we conducted two experiments. The experiments train the ResNet32 model on the CIFAR-10 dataset with an 8-worker cluster, with each configuration repeated five times (additional experiment details can be found in Section~\ref{subsec:eval_methodology}).
First, we evaluate the converged accuracy when training with different combinations of BSP and ASP protocols, as shown in Figure~\ref{fig:bsp_order}. We observe that training with BSP for 50\% of the workload and then switching to ASP outperformed its counterpart, i.e., ASP $\rightarrow$ BSP. Second, we examine the relationship between converged accuracy and the percentage of BSP training and find that training more with BSP does not necessarily lead to higher converged accuracy. As shown in Figure~\ref{fig:bsp_percentage}, the converged accuracy first increases with the percentage of BSP training and then stays on par with training entirely with BSP (i.e., 100\%). This observation also provides the basis for deriving the timing policy as described in Section~\ref{subsec:timing_policy}.

\subsubsection{Theoretical Explanations}
Next we explain why using BSP in the earlier epochs and then switching to ASP can allow us to simultaneously accomplish both our goals of minimizing the \emph{population loss} (that is, the expected value of the testing loss) as effectively as when using \emph{static BSP}, and improving the per-epoch computation time.
Here, by static BSP (respectively, ASP), we mean the protocol where one uses only BSP (respectively, ASP) from start to finish.

First, we note that the steps taken by SGD earlier in the training are much larger than those taken later in the training.  This is because, \1 at earlier epochs  the gradient tends to have a much larger magnitude, and \2 we use a decaying learning rate schedule, where the learning rate is much larger at earlier epochs than at later epochs, a common practice when training deep learning models \cite{he2016deep, huang2017densely}.

\begin{figure}[t!]
    \centering
    \begin{subfigure}[t]{0.24\textwidth}
        \centering
        \includegraphics[width=\textwidth]{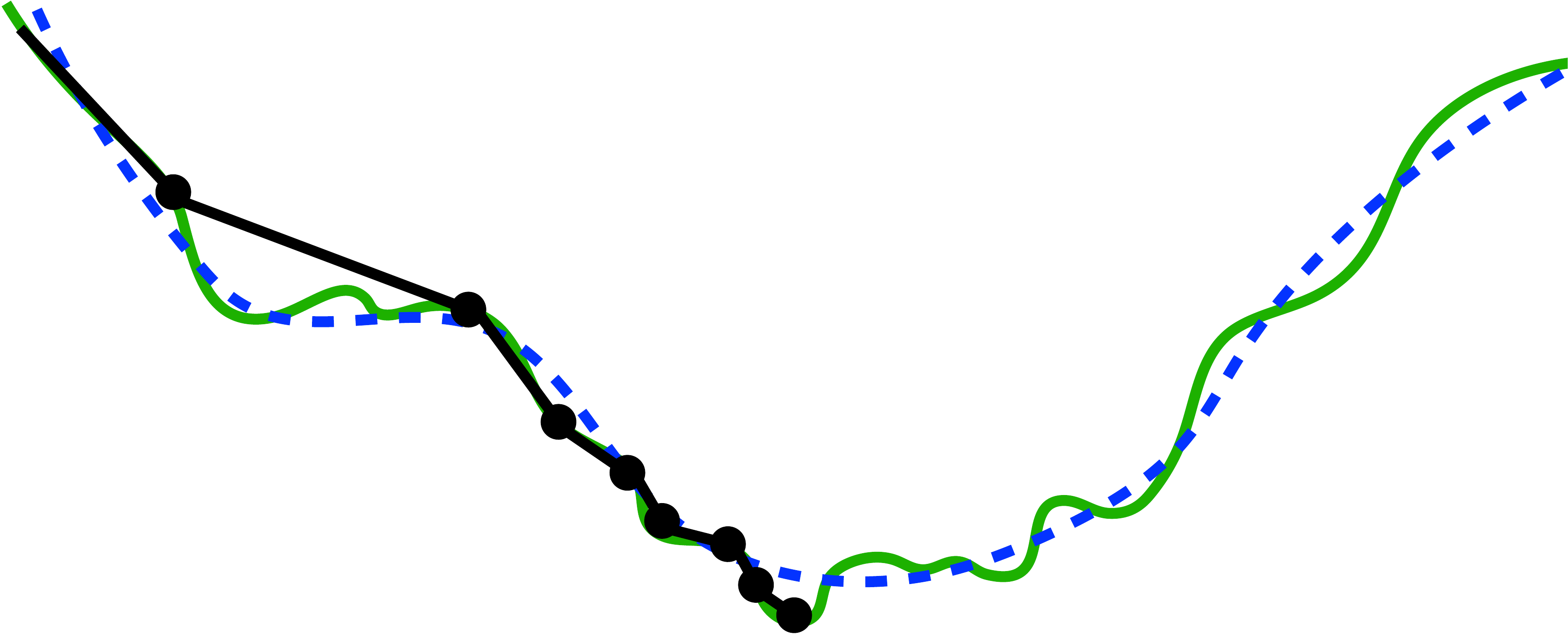}
        \caption{Static BSP.}
        \label{fig:cartoon_BSP_training_loss}
    \end{subfigure}
    \hfill
    \begin{subfigure}[t]{0.24\textwidth}
        \centering
        \includegraphics[width=\textwidth]{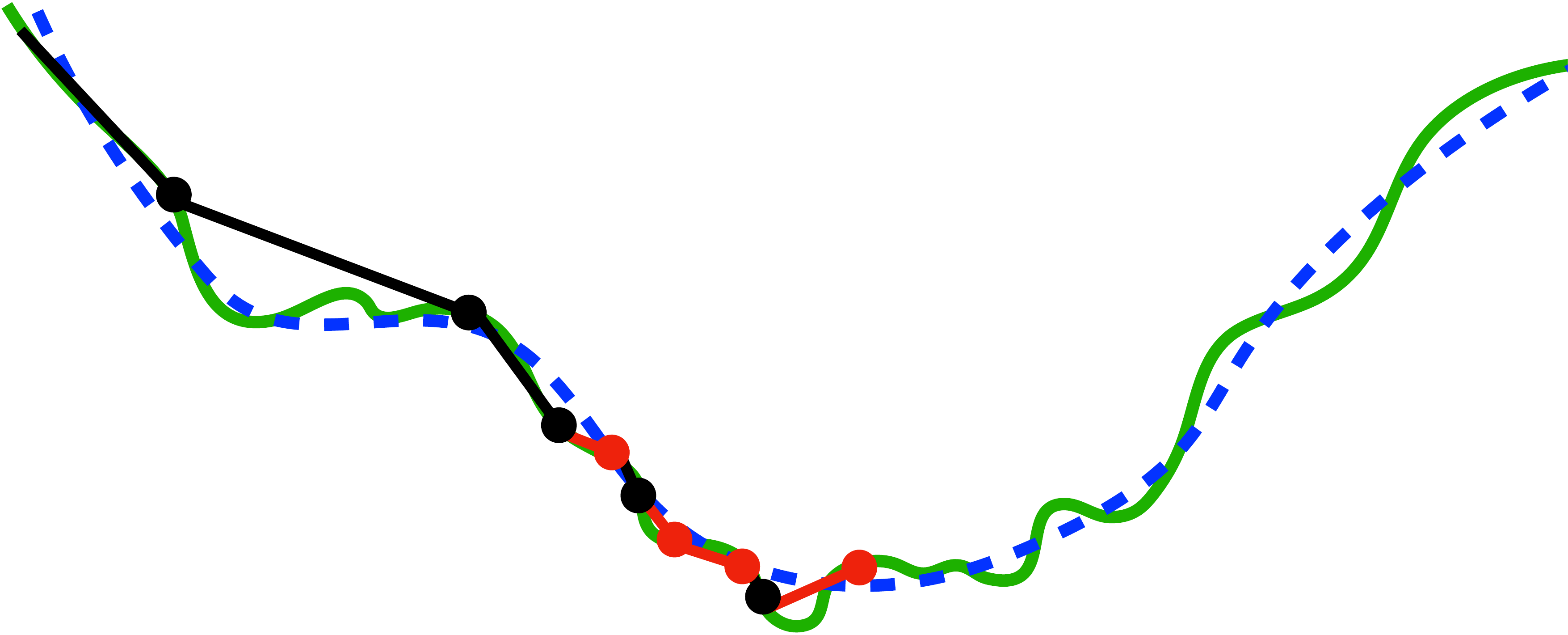}
        \caption{Switching, BSP to ASP. }
        \label{fig:cartoon_BSP_ASP_training_loss}
    \end{subfigure}
    \caption{The training loss (green solid curve) is not as smooth as the population loss  (blue dashed curve). Thus, the steps taken by \sysname (b)  with stale gradients (red lines) prevent it from finding the minimum of the training loss as effectively as static BSP (a). However, it is still able to find a point which minimizes the {\em population} loss (and hence the {\em test} loss) as effectively as static BSP.} \label{fig:cartoon_training_loss}
    \vspace{-1em}
\end{figure}

%
Second, we note that at coarser scales corresponding to the large steps taken by the algorithm earlier in the training, the landscape of the {\em population} loss  resembles that of the training loss (see Remark~\ref{rem_population_loss} in the appendix).
%
On the other hand, past empirical work~\cite{kleinberg2018alternative, hochreiter1995simplifying, keskar2016large} 
suggests that at finer scales corresponding to the smaller steps taken later in the training, the landscape of the population loss is much smoother than that of the training loss. 
 These two facts imply that while stale gradients may be ineffective very early in the training when larger steps are taken by the algorithm, stale gradients can be used to effectively minimize the population loss later in the training, since the landscape of the population loss is much smoother on a finer scale corresponding to the smaller steps taken by the algorithm later in the training (Figure \ref{fig:cartoon_training_loss}).
 This is because, at later epochs, the gradient of the population loss does not change as quickly each time the algorithm takes a step, which means that stale gradients  may still be able to provide useful information about the gradient of the population loss at the current point even though the stale gradients were computed at previous (but nearby) points.  This suggests that using ASP at later epochs (Figure \ref{fig:cartoon_BSP_ASP_training_loss}) can allow one to minimize the {\em population} loss (and hence the testing loss) as effectively as static BSP (Figure \ref{fig:cartoon_BSP_training_loss}), despite the fact that static BSP can achieve a lower {\em training} loss value (see Remark \ref{rem_can_BSP_to_ASP_minimize_training_loss} in the appendix).

\begin{figure}[t!]
    \centering
    \begin{subfigure}[t]{0.23\textwidth}
        \centering
        \includegraphics[width=\textwidth]{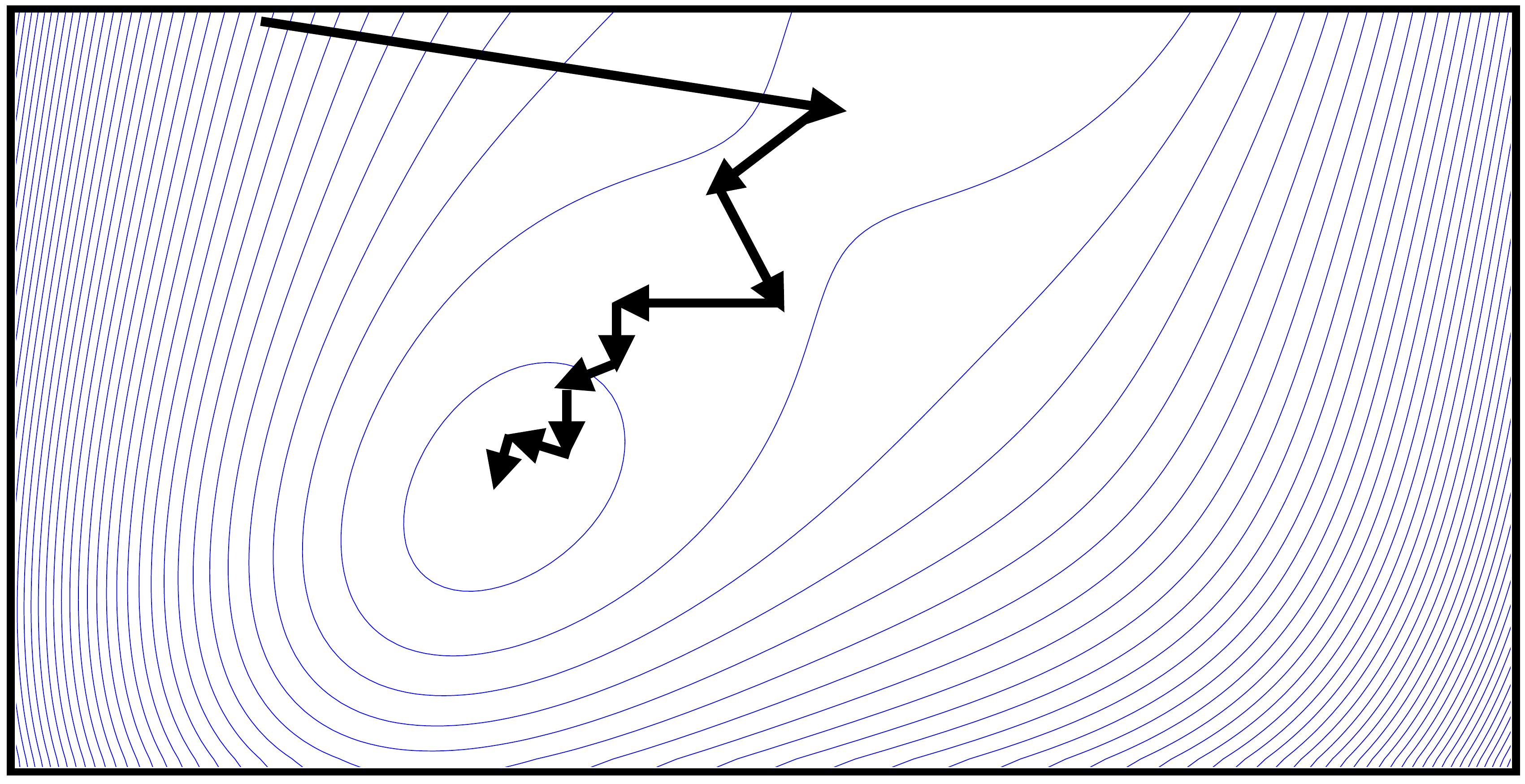}
        \caption{Static BSP.}
        \label{fig:cartoon_BSP}
    \end{subfigure}
    \hfill
    \begin{subfigure}[t]{0.23\textwidth}
        \centering
        \includegraphics[width=\textwidth]{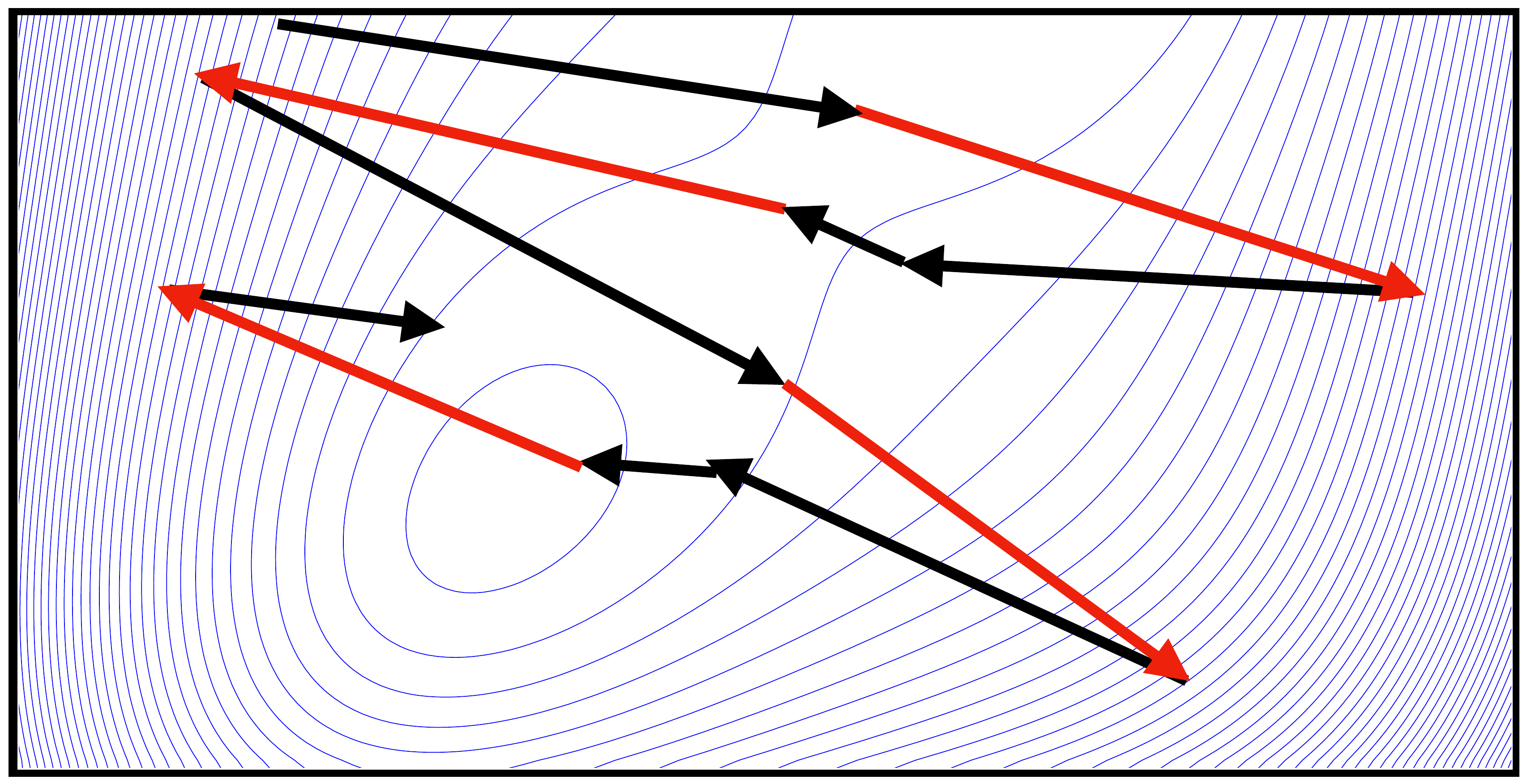}
        \caption{Static ASP.}
        \label{fig:cartoon_ASP}
    \end{subfigure}
    \hfill
    \begin{subfigure}[t]{0.23\textwidth}
        \centering
        \includegraphics[width=\textwidth]{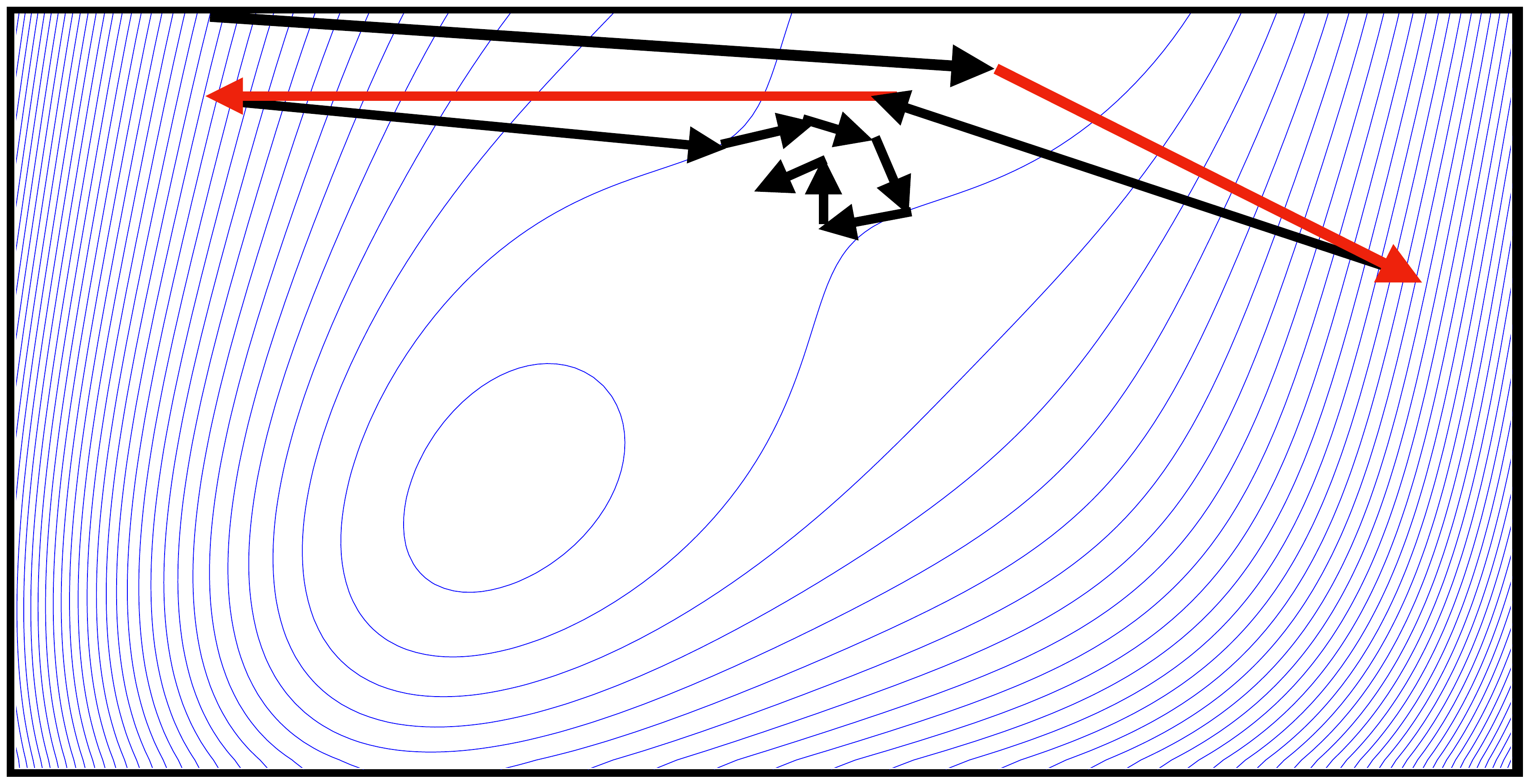}
        \caption{ASP to BSP.}
        \label{fig:cartoon_ASP_to_BSP}
    \end{subfigure}
    \hfill
    \begin{subfigure}[t]{0.23\textwidth}
    \centering
    \includegraphics[width=\textwidth]{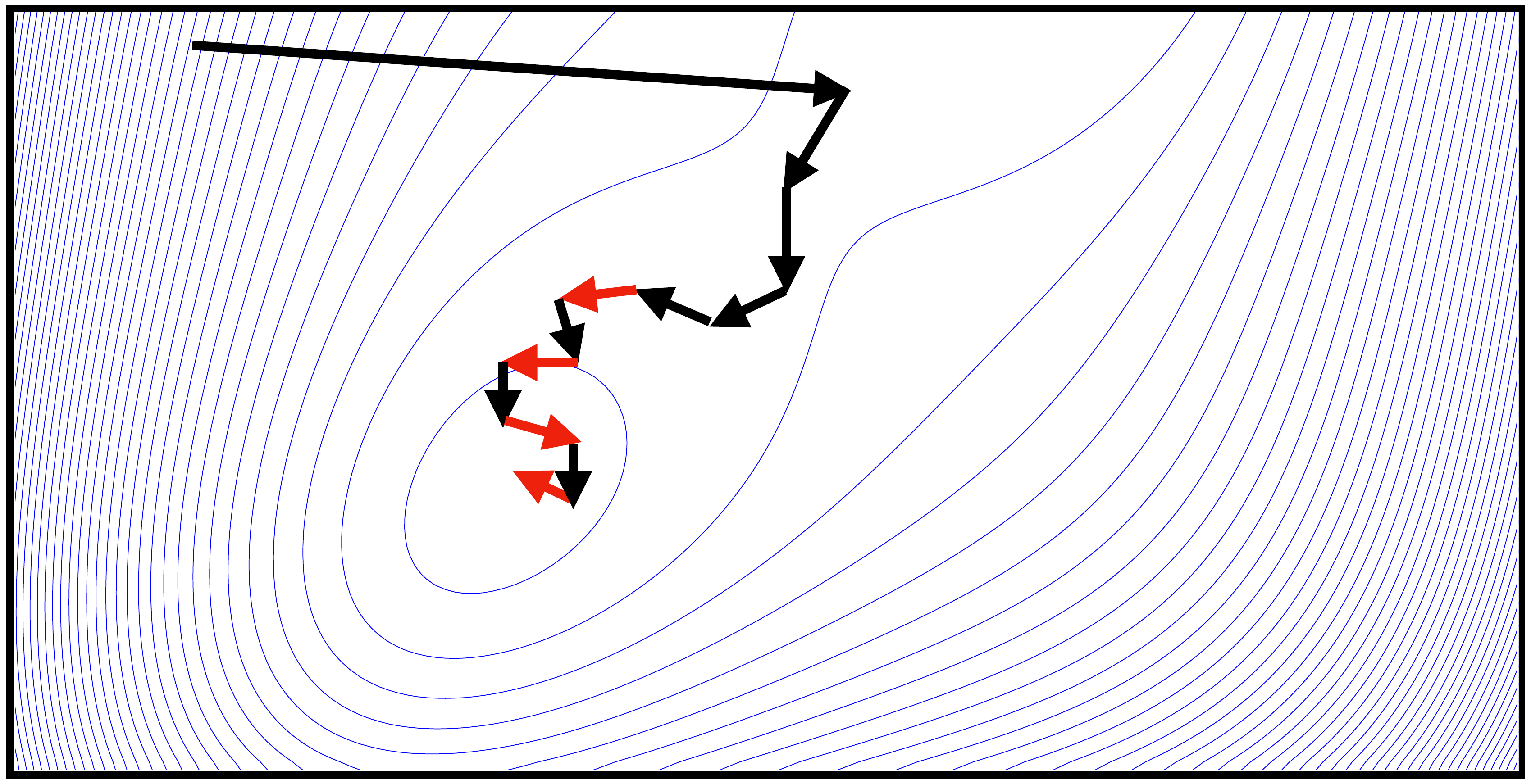}
    \caption{BSP to ASP.}
    \label{fig:cartoon_BSP_to_ASP}
    \end{subfigure}
    \caption{\textbf{Illustrations of various combinations of BSP and ASP on a simple smooth loss function (blue level sets).} 
    In earlier epochs, the gradient changes quickly at each step and stale gradients (red) are much less reliable than non-stale gradients (black).  Thus algorithms which use ASP earlier in the training may have a difficult time settling into a local minimum region where the gradient is small (b, c). This is true even if one switches from ASP to BSP (c) as the algorithm may get stuck in a saddle point once the learning rate has decayed, causing it to take a long time to escape the saddle point.  In contrast, if one only uses stale gradients at later epochs once the algorithm is closer to a minimum point, the stale gradients will be low-bias estimates for the true gradient and will still allow the algorithm to reach the minimum point (d).
    %
    %
    \label{fig:cartoon_switchin_orders}}
    
    \vspace{-1em}
\end{figure}

On the other hand, at very early epochs, when the algorithm is very far from a minimum point, the gradient of the loss function may have a very large magnitude and may change very quickly at each step \footnote{For instance, consider the simple 4'th-order polynomial $f(x) = x^4$.  For this function, the gradient $f'(x) = x^3$ changes more quickly when $x$ is further from the minimum point $x=0$.  Roughly speaking, for loss functions defined by deep neural networks there can be many (multivariate) high-order terms, with the order of the terms growing with the depth of the network.} 
Thus, in protocol policies where ASP is used early in the training, its stale gradients may not provide much information about the value of the gradient at the current point, and may cause ASP to exhibit unstable behavior early in the training (Figures \ref{fig:cartoon_ASP} and \ref{fig:cartoon_ASP_to_BSP}), preventing ASP from effectively decreasing the loss value. 
To avoid this unstable behavior, one should use BSP in the very early stages of the training, and only switch to ASP at a later epoch once the learning rate is lower and the gradients are changing more slowly (Figure \ref{fig:cartoon_BSP_to_ASP})
(See Remark \ref{rem_why_not_ASP_to_BSP} in the appendix.)

\subsection{Timing Policy: When to Switch?} 
\label{subsec:timing_policy}


Next, we introduce \sysname's timing policy which determines when to switch between BSP and ASP. Deciding the proper timing to switch is an important problem as it not only impacts the converged accuracy but also the training time. Furthermore, it is a challenging problem as both model-specific factors such as training progress and runtime factors such as slow nodes can impact the timing. Specifically, we develop both offline and online policies that are suitable to use under different training conditions. 

\subsubsection{Offline Policy via Binary Search}
\label{subsec:offline_binary_search}


Based on our empirical observation that BSP is strictly slower than ASP, with or without stragglers, and that the converge accuracy increases monotonically with the amount of BSP training, we formulate the searching process as a binary search problem.
%
Specifically, for a given training workload, our goal is to find a switching point $s$ that yields a converged accuracy $\alpha(s)$ that satisfies $\alpha(s_{min}) \leq \alpha(s) \pm \alpha_{threshold} \leq \alpha(s_{max})$, where $s_{min}$ and $s_{max}$ represents training with ASP and BSP, respectively.
Further, the corresponding training time $T(s)$ should be as close to $T(s_{min})$ as possible, i.e., switching as early from BSP to ASP as possible. Lastly, we want to find $s$ in as fewer trial training sessions as possible to reduce search overhead. 
The pseudo-code for our binary search algorithm~\ref{appendix:sec:pseudo} can be found in the appendix. 
%
%
In summary, for each distributed training workload, we can use a binary search algorithm to find the best switching point $s$ at which point \sysname will switch from training with BSP to ASP. We analyze the cost and performance tradeoff in Section~\ref{subsec:search_overhead}.

\subsubsection{Online Policies for Handling Stragglers}
\label{subsubsec:online_policies}

The offline policy described above provides us a good basis for speeding up distributed training while achieving high-quality test accuracy. However, it does not account for runtime factors including stragglers and also requires upfront search cost. In this section, we introduce two types of policies that are designed to react to the training status. 


We target transient stragglers, e.g., nodes that exhibit temporary slowdown due to datacenter network or server resource contention, a non-rare occurrence when using public cloud~\cite{duan2017cloud,Li2010-dm}. 
Note that permanent stragglers are best dealt with by requesting replacement~\cite{Or2020-nh,Peng2018-mw}---simply switching to a more straggler-tolerant protocol like ASP might mitigate but not eradicate the performance impact. 
More explicitly, we consider the policies that deal with permanent stragglers as complementary work and therefore skip the discussion and evaluation of such policies. 
For ease of exposition, we assume that each occurrence of the slowness lasts at most the time to provision a new cloud server---we use 100 seconds based on empirical measurement reported by prior work~\cite{Li2020-bk}.
We further assume that \1 each node can become a straggler at any time during the training; \2 the number of unique straggler nodes is less than the cluster size.  
Our goal is to design policies that adequately deal with the potential impact on the training time which also work \emph{in tandem} with the other policies of \sysname described above. 

We introduce two policies, both centering on the key insight that any transient straggler-oriented policies only need to react before the switch timing for a given workload. This is because once a training session is switched to ASP, we consider it immune from the impact of transient stragglers. 
The first \emph{greedy} policy simply switches to ASP (if it is not already using ASP, which can happen if two stragglers overlap) when a straggler is detected; once the cluster is free of any stragglers and the aggregate BSP training has not been satisfied, it will switch back to training with BSP. This policy therefore might result in multiple synchronization switches, which overhead as we will show in Section~\ref{subsec:practical_overhead} can be in the order of tens seconds. 

To circumvent the switching overhead, we design an elastic-based policy that removes any detected stragglers from the current cluster so as to complete the specified amount of BSP training free of stragglers. Once the designated BSP workload is fulfilled, it will then restore the cluster size and train the remaining workload with ASP. As such, this elastic-based policy is more resistant to the frequency of straggler occurrences.
For both policies, we leverage the historical average training throughput to detect the stragglers, a common technique used in various application domains~\cite{Li2020-bk, xie2010improving}. 
Specifically, a worker $k$ is identified as a straggler if its training throughput over a sliding window $S_k$ is lower than the difference between the cluster average and standard deviation $S-\sigma$, for a number of consecutive detection windows.

\subsection{Configuration Policy: How to Adjust Hyper-parameters?}
\label{subsec:hyperparameter}

\begin{figure}[t]
    \centering
    \begin{subfigure}[t]{0.2\textwidth}
        \includegraphics[width=\textwidth]{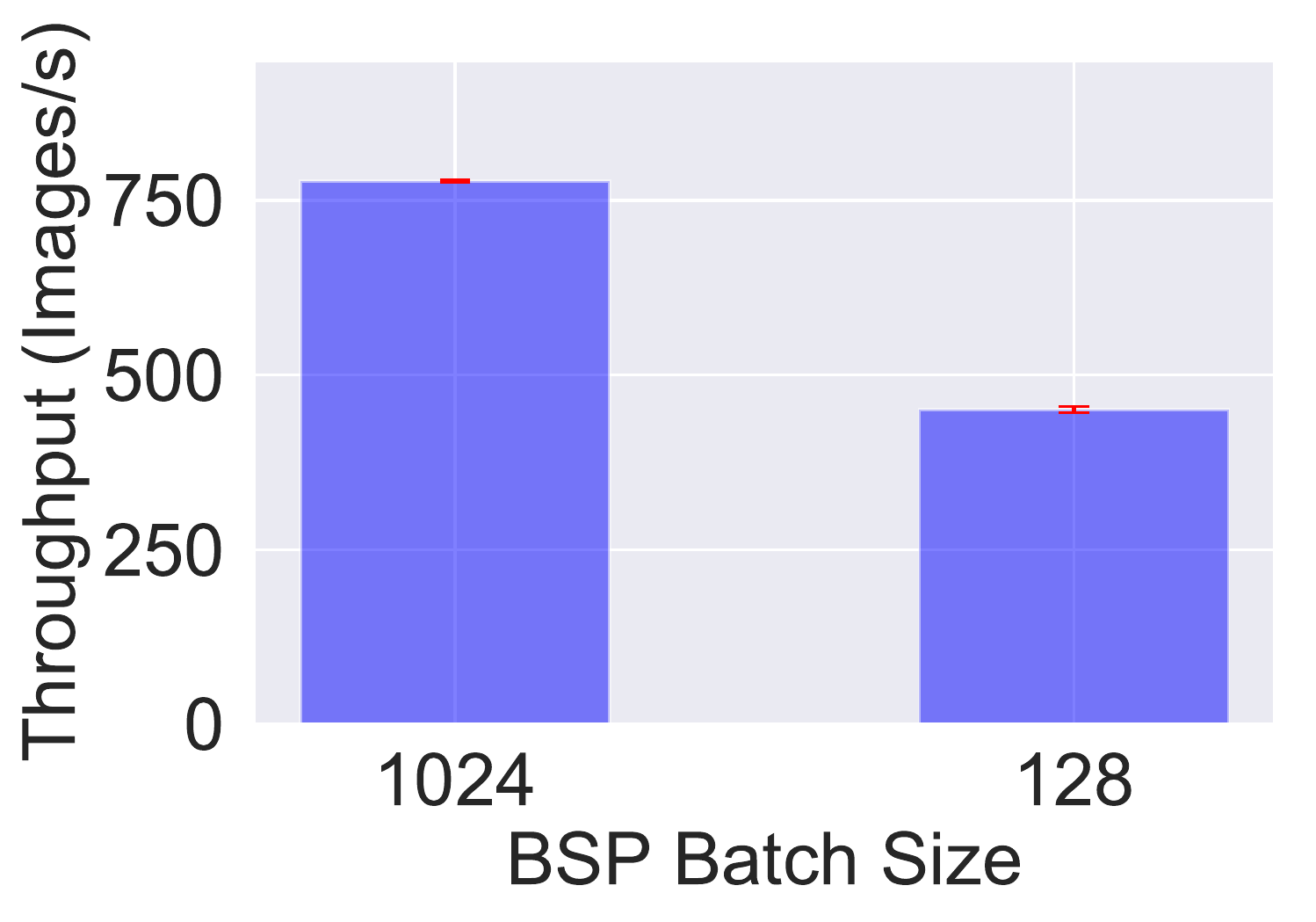} 
        \caption{Batch size scaling.}
        \label{fig:batch_size_scaling}
    \end{subfigure}
    \hfill
    \begin{subfigure}[t]{0.28\textwidth}
        \includegraphics[width=\textwidth]{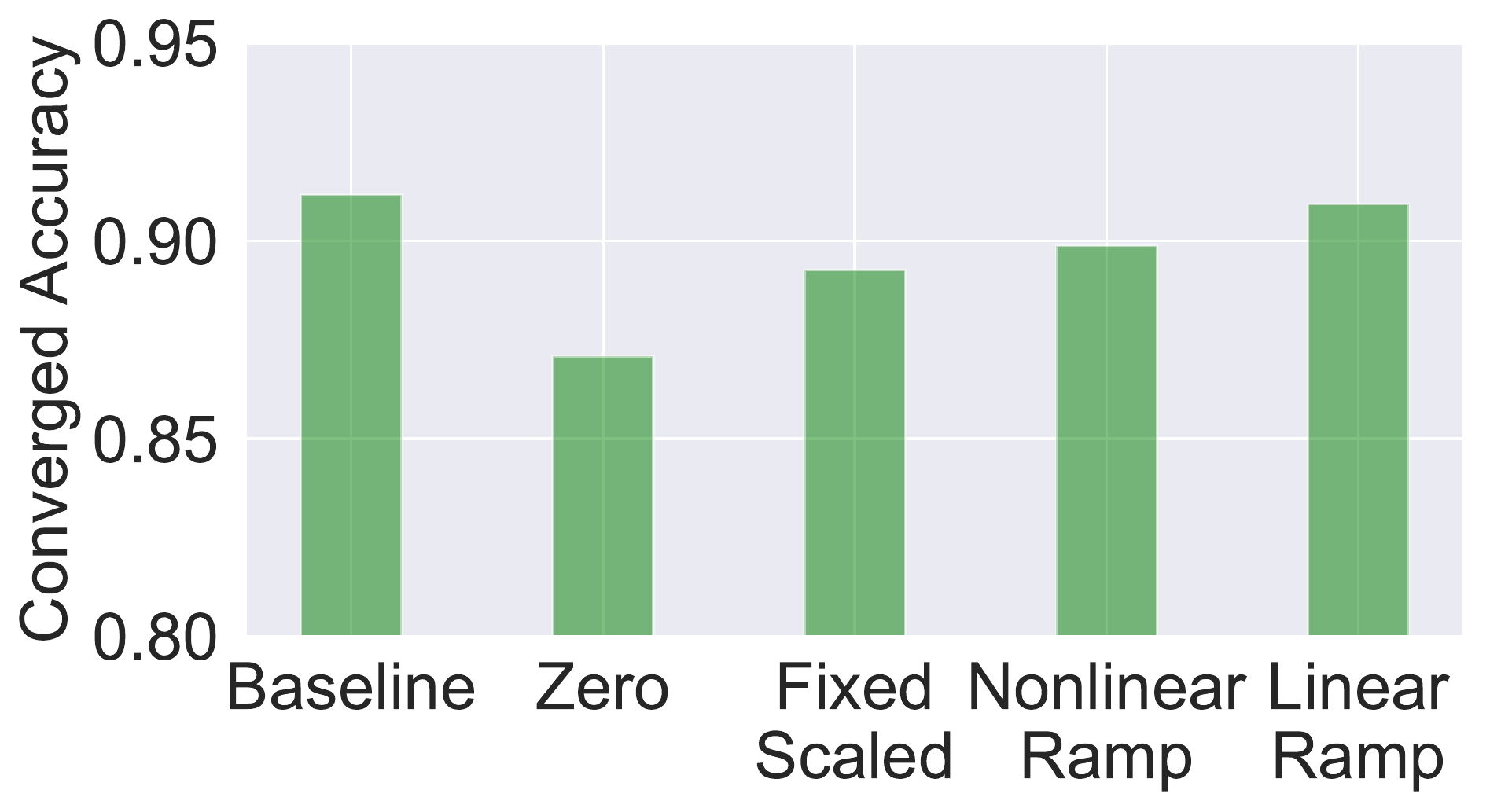} 
        \caption{Momentum Scaling.
        }
        \label{fig:momentum_scaling}
    \end{subfigure}
    \caption{\textbf{Comparison of different hyper-parameter configurations: Exp. Setup 1.} 
    The four methods we explored for setting momentum after switching to ASP are: \1 setting the momentum to 0, \2 setting the momentum to $\frac{1}{n}$, \3 ramping up the momentum based on $\frac{2^i}{n}$, \4 ramping up the momentum based on $\frac{i}{n}$ where $i$ is the number of epochs after switching. Both \3 and \4 will stop the ramp up once the momentum reaches the original value used by BSP.
    }
    \label{fig:hparam_config}
    \vspace{-1em}
\end{figure}

Hyper-parameters are commonly considered important for training performance~\cite{senior2013empirical, smith2017don}. In our work, we also observe non-negligible differences with different hyper-parameters \emph{after switching} from BSP to ASP.
For example, in Figure~\ref{fig:batch_size_scaling}, we show that the difference in training throughput can be up to 2X with different batch sizes; in Figure~\ref{fig:momentum_scaling}, we observe up to 5\% converged accuracy differences using different momentum scaling techniques. 
As such, it is important to choose suitable hyper-parameters after synchronization switching so as to retain the training benefits. 

Through empirical investigations, we find that adjusting the following three hyper-parameters: \emph{mini-batch size}, \emph{learning rate}, and \emph{momentum} sufficient. 
Note, since our work is not about hyper-parameter tuning, we do not focus on finding the optimal hyper-parameters set for a specific training setting; rather, we are aiming to automatically change the value of these hyper-parameters based on the synchronization protocol in use. 
We assume the deep learning practitioners will provide an initial set of hyper-parameters, e.g., a mini-batch size of $B$ and a learning rate of $\eta$, given the training workload and a GPU cluster of $n$ nodes. 

To start the training with BSP, we will configure the mini-batch to be $nB$. This configuration is based on both the implementation of BSP batch size in the TensorFlow framework (i.e., as a global value distributed to each worker) and prior work's suggestion of setting BSP batch size proportionally to the cluster size~\cite{goyal2017accurate}. Once switching from BSP to ASP, we will reduce the mini-batch to be $B$ as in ASP training the batch size is treated as a local value specific to each worker. 
We use the \emph{linear scaling rule} for setting the learning rate for training with BSP as $\eta_{BSP} = n\eta$. This is based on prior work that demonstrated the effectiveness of scaling learning rate based on mini-batch size~\cite{goyal2017accurate}.
In contrast to prior work that adjusts momentum based on mini-batch size~\cite{lin2019dynamic}, we find that using the \emph{same} momentum value for both BSP and ASP allows \sysname achieve comparable accuracy to training with BSP (i.e., leftmost bar in Figure~\ref{fig:momentum_scaling}).

\section{\sysname Implementation}
\label{sec:impl}

\begin{figure}[t]
    \centering
        \includegraphics[width=0.48\textwidth]{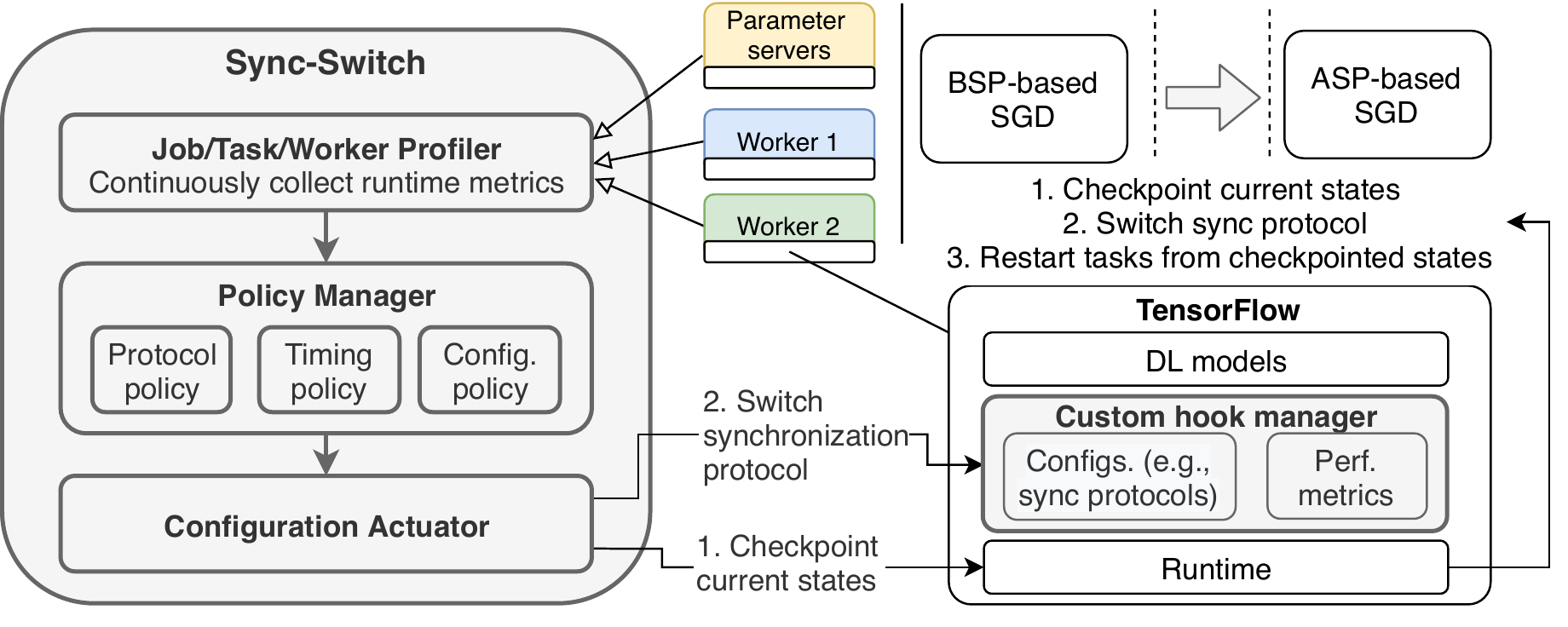}
        \caption{
        \textbf{\sysname architecture and implementation.} We implement \sysname on top of the popular TensorFlow framework as two logical parts: a standalone entity and a per-node part. Grey-shaded components are our modifications. Hollow and solid arrows represent the profiling and actuation workflow, respectively. 
        } 
        \label{fig:sgd_switcher}
\end{figure}

We implemented a \sysname prototype, as shown in Figure~\ref{fig:sgd_switcher}, based on TensorFlow v1.10 and Tensor2Tensor v1.9~\cite{tensor2tensor}. The prototype includes the parameter synchronization policies described in Section~\ref{sec:design} for distributed training jobs. \sysname users can manage their distributed training jobs via the command line. \sysname's implementation consists of two logical components: a standalone cluster manager that interfaces with Google Cloud Platform
and a custom hook manager embedded in TensorFlow that collects training status and adjusts per-node configurations.

The \emph{cluster manager} first takes the user input, including the training job script and cluster size, to initialize protocol and configuration policies.
If the job is a recurring one, the cluster manager initializes the timing policy based on the prior binary search-based result. 
Otherwise, the cluster manager launches a pre-specified number of pilot jobs per the search algorithm described in Section~\ref{subsec:offline_binary_search} to obtain the timing policy. 
Afterward, \sysname creates the training cluster consisting of nodes running with TensorFlow and sets up the \emph{profiler} for continuously collecting runtime metrics.

\sysname's custom hook manager is written as a core Python component to interact with TensorFlow runtime to collect internal metrics such as training throughput and training loss and to change hyper-parameters like learning rates. 
The collected metrics are sent back to the profiler, which, in conjunction with the policy manager, decides whether to trigger a synchronization protocol switch. 
The switch mechanism is implemented by having each custom hook manager listen at a pre-specified port for incoming commands and by leveraging TensorFlow's built-in model checkpoint/restore functions for persisting the training progress. In \sysname, once all custom hook managers finish checkpointing, the cluster manager propagates the updated training job and configurations to all nodes. Once notified, custom hook managers relaunch the training tasks to resume the training from the last model checkpoint but with a different synchronization protocol.  

\section{Evaluation}
\label{sec:eval}

\begin{table*}[t]
\centering
\footnotesize
\begin{tabular}{r|l|c|l||cc|cc} 
\toprule
\multicolumn{1}{c|}{\textbf{Experiment}\textbf{~}~} & \multicolumn{1}{c|}{\textbf{Workload}} & \textbf{Cluster} & \multicolumn{1}{c||}{\textbf{\sysname policy}} & \multicolumn{2}{c|}{\textbf{Throughput Speedup}} & \multicolumn{2}{c}{\textbf{TTA Speedup}}  \\
\multicolumn{1}{l|}{\textbf{Setup}}           & \multicolumn{1}{c|}{(model, dataset)}  & (size, GPU)              & \multicolumn{1}{c||}{(protocol, timing)} & vs. ASP & vs. BSP                       & vs. ASP & vs. BSP                 \\ 
\midrule
1                                    & ResNet32, CIFAR-10                      & 8, K80                   & P1: ([BSP, ASP], 6.25\%)                       & 0.78X   & 5.13X                         & N/A      & 3.99X                 \\
2                                    & ResNet50, CIFAR-100                     & 8, K80                   & P2: ([BSP, ASP], 12.5\%)                       & 0.89X   & 1.66X                         & N/A      & 1.60X                   \\
3                                    & ResNet32, CIFAR-10                      & 16, K80                  & P3: ([BSP, ASP], 50\%)                         &  failed      & 1.87X                         & N/A      & 1.08X~                  \\
\bottomrule
\end{tabular}
\caption{\textbf{Summary of our experiment setups, timing policies, and performance.} We observe that \sysname achieves up to 5X (4X) speedup in training throughput (TTA). 
$P_i$ represents the set of policies for setup $i$. Note ASP-based training failed in exp. setup 3. 
}
\vspace{-1em}
\label{table:exp_summary_policies}
\end{table*}

We conducted our experiments by training popular deep learning models on Google Cloud Platform (GCP) to quantify \sysname's performance over training exclusively with BSP and with ASP, two commonly chosen baselines~\cite{jiang2019novel,dutta2020slow}. 
%
Note, since the performance of existing synchronization protocols all fall in between that of BSP and ASP, we believe evaluating using BSP and ASP provide us a good foundation for understanding \sysname's performance. 
Furthermore, semi-synchronous protocols, such as SSP and DSSP, can also be utilized in \sysname (for example switching from SSP to ASP)---\sysname is agnostic to the underlying synchronization protocols. 
%
Our evaluation includes systems experiments using our \sysname prototype to evaluate the efficacy of timing policies
and framework overhead with real TensorFlow jobs, as well as simulation experiments to analyze the performance and cost of our binary search-based algorithm  under realistic workload conditions. 
Table~\ref{table:exp_summary_policies} summarizes our experiment setups and result highlights. 

\subsection{Evaluation Setup and Methodology} 
\label{subsec:eval_methodology}

\para{Distributed Training Workloads.} We choose two different workloads, \1 ResNet50 on the CIFAR-100 dataset and \2 ResNet32 on the CIFAR-10 dataset.
Both models are part of the ResNet model family, one of the widely used CNNs for image recognition tasks. We use the ResNet implementations from Tensor2Tensor~\cite{tensor2tensor}.
ResNet50 has more layers than ResNet32, leading to different model parameter size and floating-point operations, and therefore has longer per-batch 
training time with the same cluster.  
The datasets, each containing 60K images of size 32X32 pixels, are widely used in deep learning research~\cite{cifar10}. The key difference between the two datasets is the number of the classification classes (i.e., CIFAR-100 contains 100 classes vs. CIFAR-10 contains 10 classes); therefore, it often takes more epochs to train on the CIFAR-100. 
As such, these two workloads allow us to evaluate \sysname's performance under different computation and learning requirements.

\para{Cluster Setup and Configuration.} 
We run all experiments on cloud-based GPU clusters in GCP's \textit{us-west1} region; we choose two commonly used cluster sizes of eight and sixteen\footnote{Smaller cluster size has less impact on ASP's converged accuracy~\cite{Li2020-bk}.} to evaluate \sysname's performance~\cite{Or2020-nh,dutta2020slow}.
Each server, running Ubuntu 18.04 LTS, has 8 vCPUs, 30 GB of main memory, 100GB local HDD storage, and is equipped with one Nvidia K80 GPU card. 
To account for the inherent accuracy variations in SGD-based deep learning training, we repeat each experiment setup five time using the same model parameter initialization algorithm.
We report both the average performance with standard deviation and the runs with the best performance, measured by the highest achieved test accuracy.

\para{Evaluation Metrics.}
We use two groups of metrics for evaluating \sysname's efficiency in parameter synchronization (first group) and its associated overhead (second group).
The first group includes training loss, test accuracy, total training time, and time-to-accuracy. %
\emph{Training loss} is calculated based on the cross-entropy loss function per mini-batch. We report the average training loss collected every 100 ASP steps to avoid incurring excessive measurement overhead~\cite{Li2020-bk}. 
%
\emph{Test accuracy} refers to the top-1 accuracy of the trained model when evaluating on the test dataset. We measure the test accuracy every 2000 ASP steps on the standalone cluster manager to avoid impacting the training performance. 
A model is said to be converged if its test accuracy has not changed for more than 0.1\% for five evaluations
and we report the corresponding value as the \emph{converged accuracy}.
%
\emph{Total training time} is the time, including computation and networking time, taken for a training cluster to complete a user-specified workload. We measure this time at the end of each training using the TensorFlow built-in logs. 
%
\emph{Time-to-accuracy (TTA)} denotes the time to reach a specified test accuracy threshold and provides valuable insights into both the training throughput and model accuracy~\cite{coleman2019analysis}. We use the average converged accuracy of models trained with BSP in the same setup as the threshold.
For the second group, \emph{search time} quantifies the time for \sysname to find the near-optimal switch timing using the binary search. We calculate search time as the sum of the total training time of all sessions trained during the search. \emph{Switch overhead} describes the total time to switch between synchronization protocols with TensorFlow.

\para{Hyper-parameter Setting.} We set the initial hyper-parameters based on the original ResNet paper~\cite{he2016deep} and use the SGD with momentum of 0.9 as the optimizer. Specifically, we configure the training workload to be 64K steps, batch size to be 128, and learning rate to be 0.1. We use a piece-wise function that decays learning rate at 32K and 48K steps, with scaling factors of 0.1 and 0.01, respectively. Further, we use the configuration policies described in Section~\ref{subsec:hyperparameter} to adjust relevant hyper-parameters based on the specific experiment setup.

\subsection{Performance of \sysname}
\label{subsec:perf_search_based_policy}

\begin{figure}[t]
    \centering
    \begin{subfigure}[t]{0.24\textwidth}
        \includegraphics[width=\textwidth]{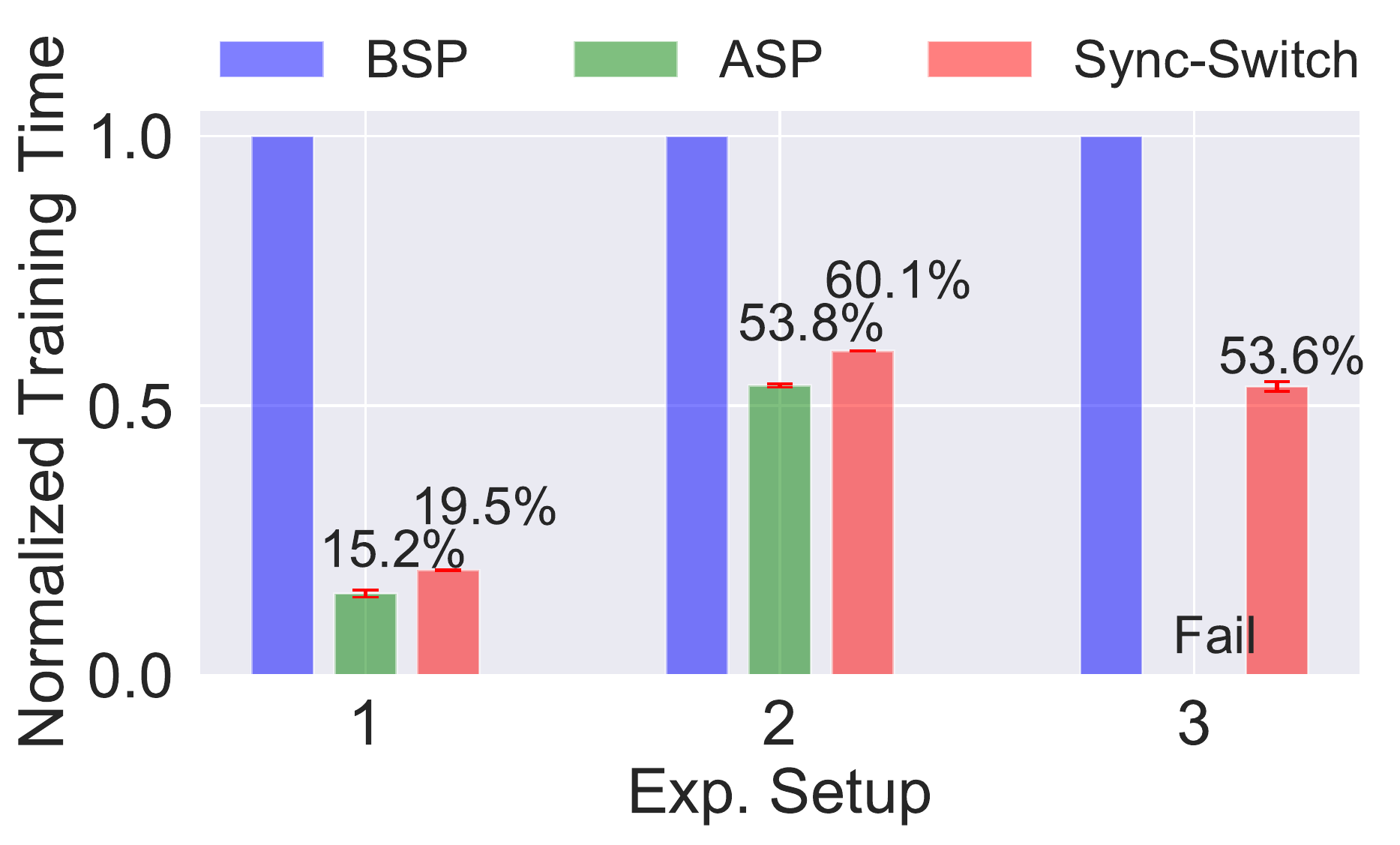}
        \caption{Total training time.}
        \label{fig:optimal_comp_time}
    \end{subfigure}
    \hfill
    \begin{subfigure}[t]{0.24\textwidth}
        \includegraphics[width=\textwidth]{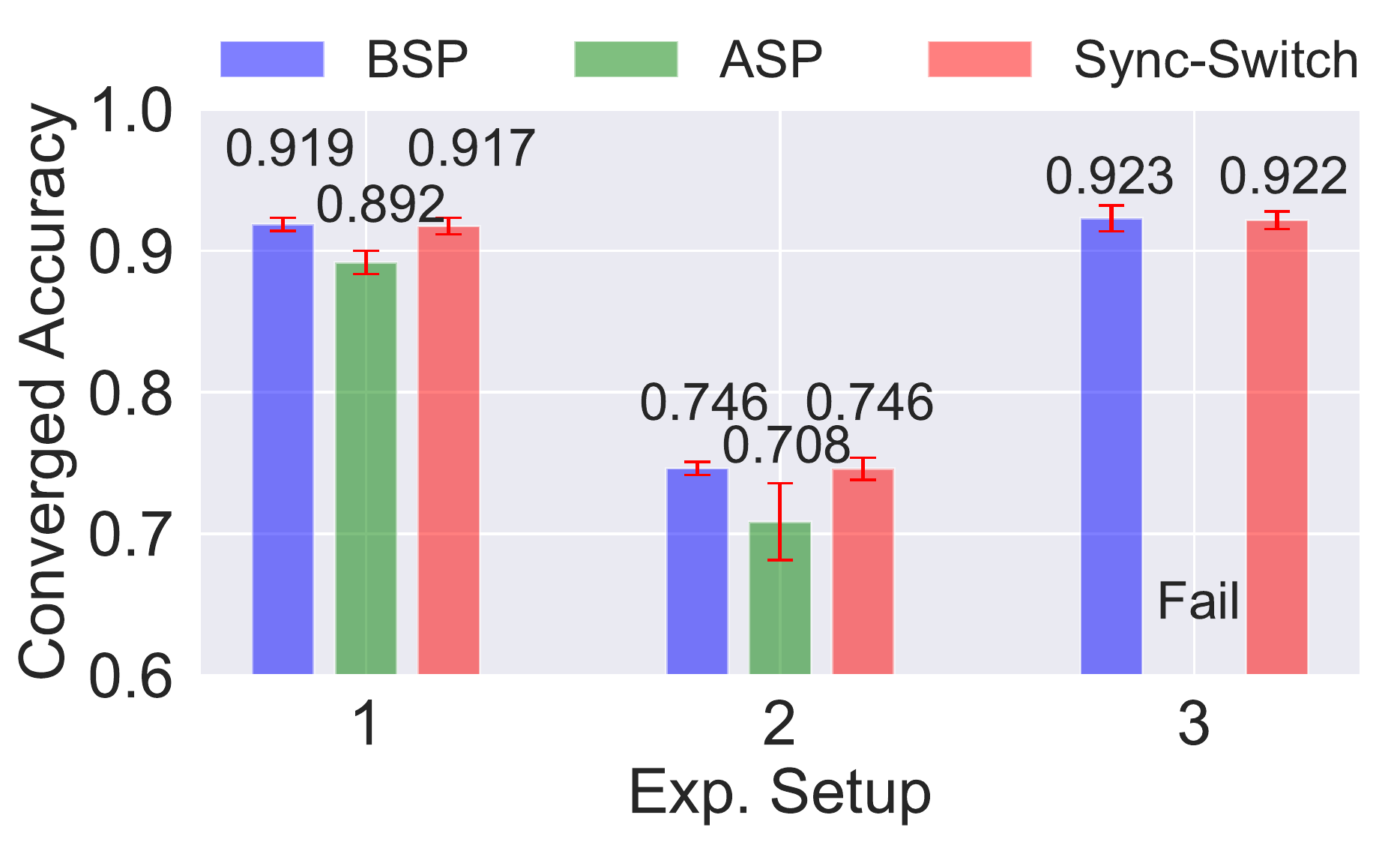}
        \caption{Converged accuracy.}
        \label{fig:optimal_comp_acc}
    \end{subfigure}
    \caption{\textbf{End-to-end performance comparison.} Using \sysname, we observe on average 1.66X-5.13X speedup and similar converged accuracy compared to training with BSP. 
    \sysname achieved up to 3.8\% higher converged accuracy compared to training with ASP. 
    %
    }
    \label{fig:eval_search_policy_overview}
\end{figure}

\begin{figure*}[t]
    \centering
    \begin{subfigure}[t]{0.24\textwidth}
        \includegraphics[width=\textwidth]{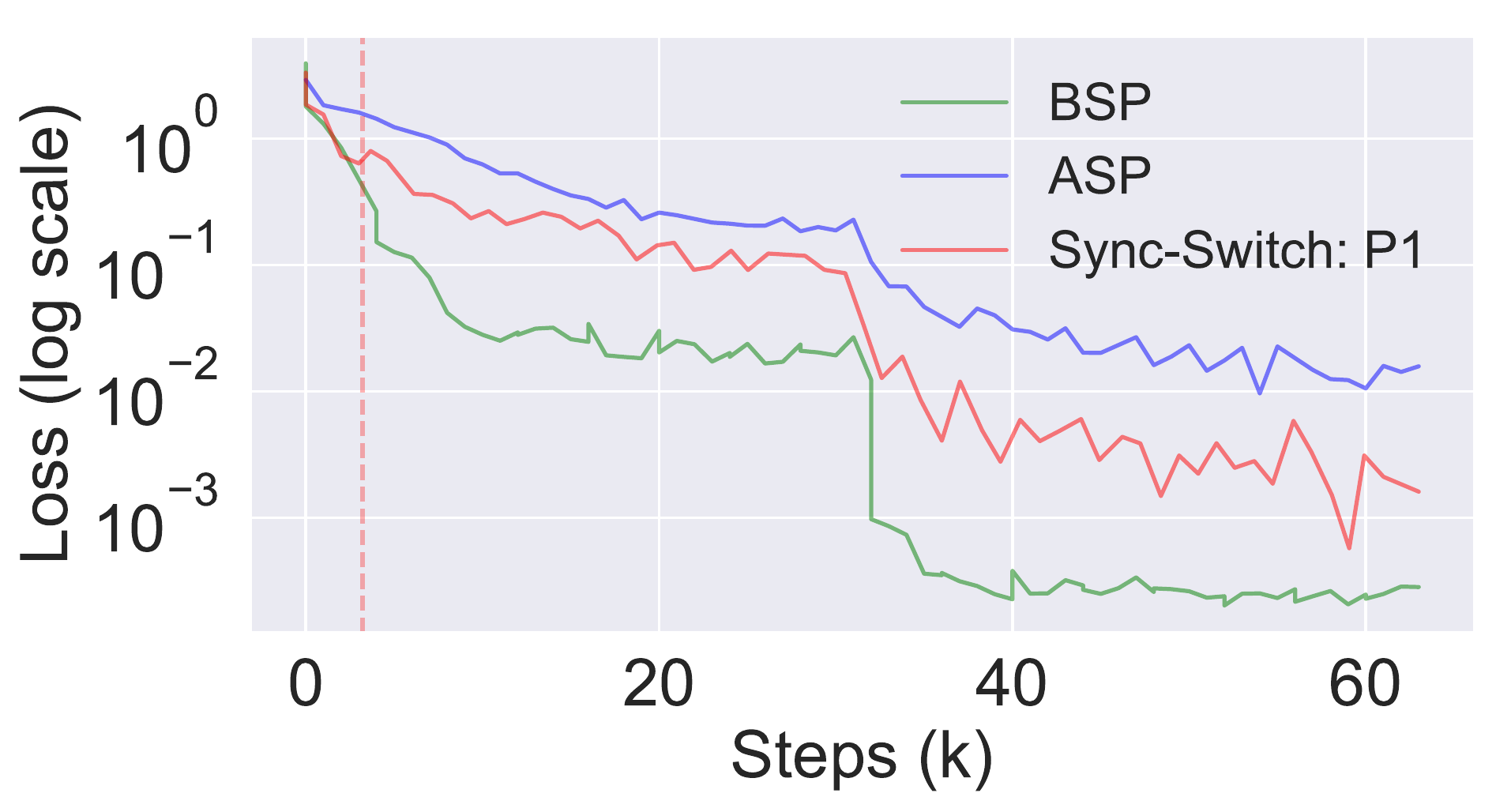}
        \caption{Training loss.}
        \label{fig:res32_cluster8_anytime_loss}
    \end{subfigure}
    \hfill
    \begin{subfigure}[t]{0.24\textwidth}
        \includegraphics[width=\textwidth]{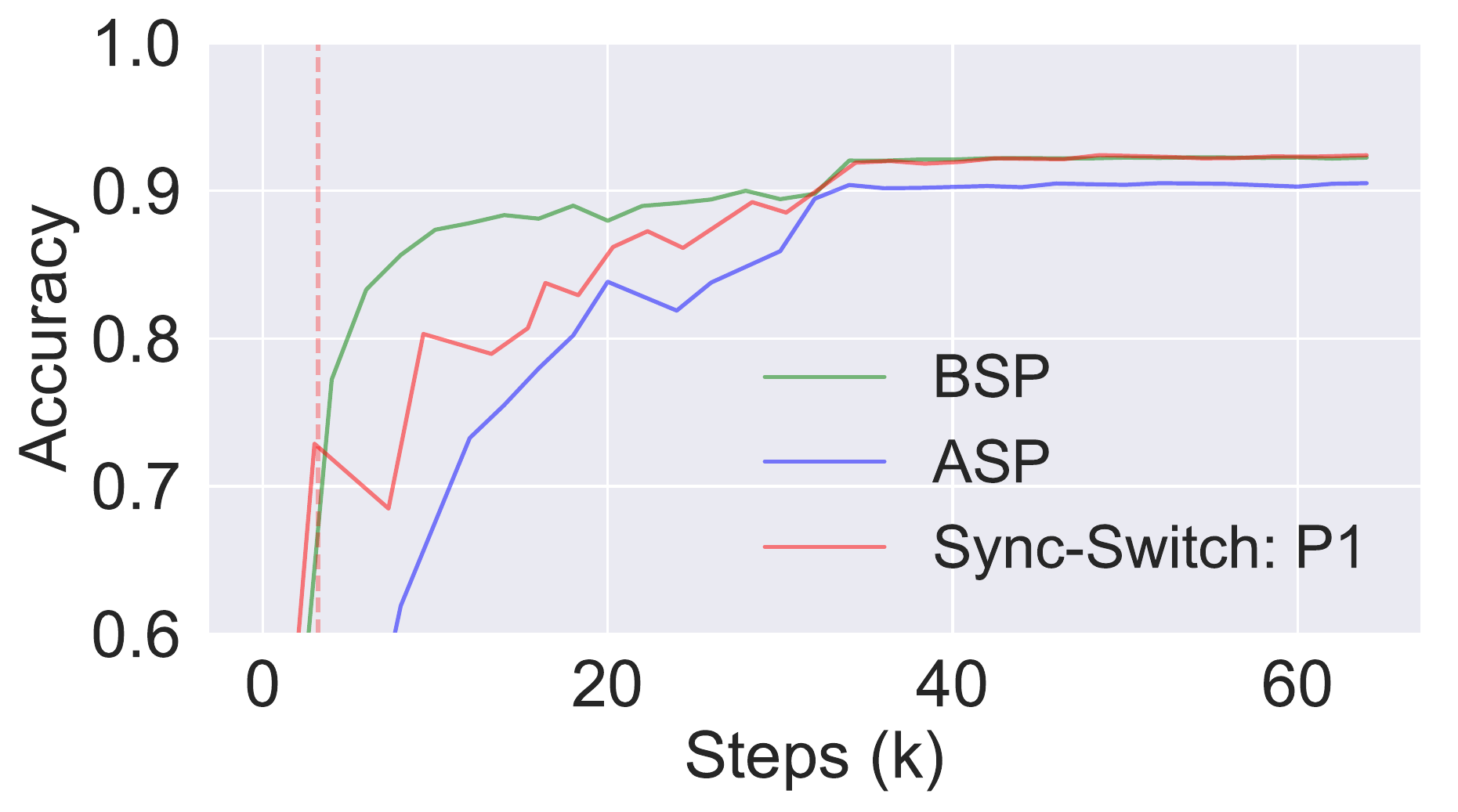}
        \caption{Test accuracy.}
        \label{fig:res32_cluster8_anytime_acc}
    \end{subfigure}
    \hfill
    \begin{subfigure}[t]{0.24\textwidth}
        \includegraphics[width=\textwidth]{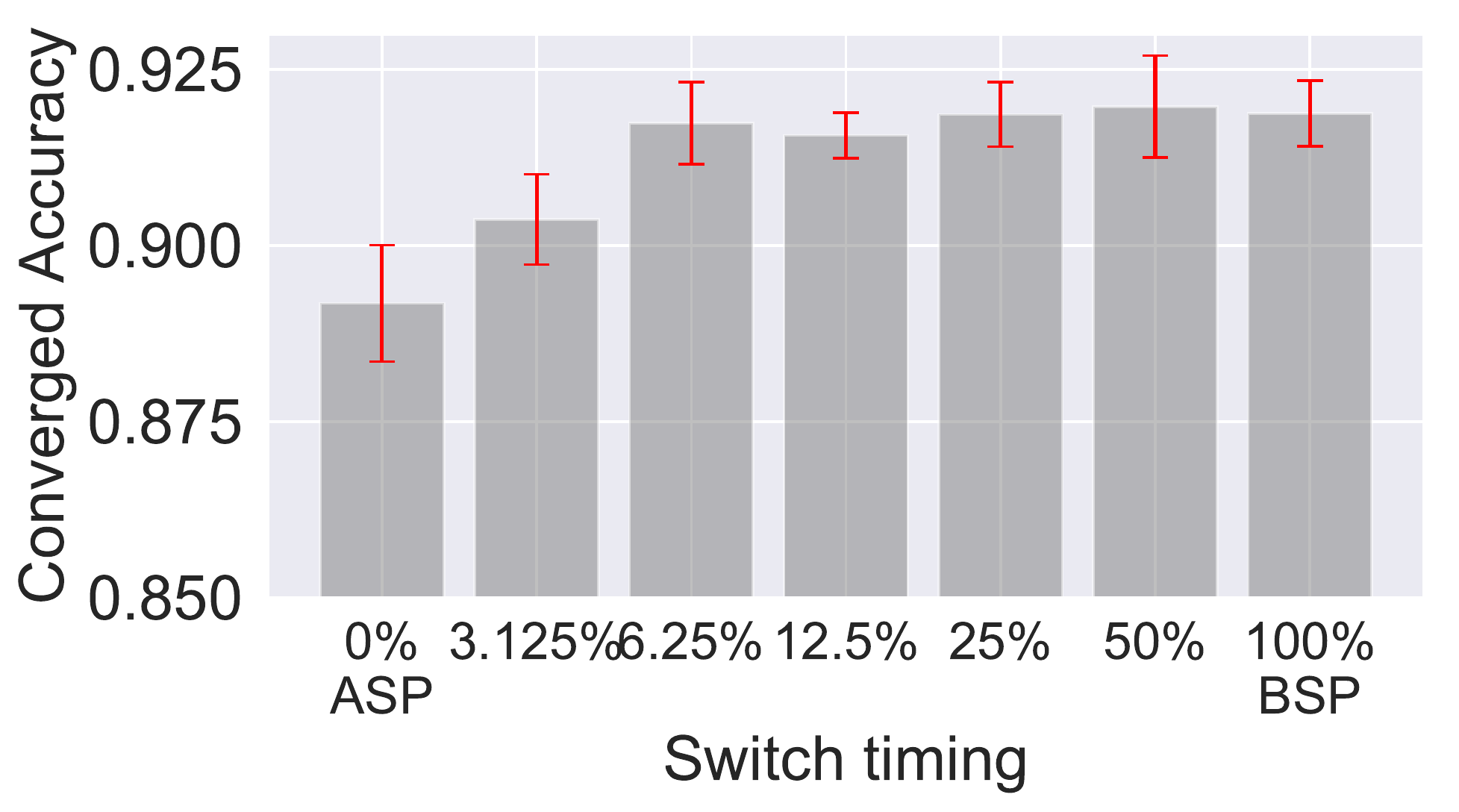}
        \caption{Converged accuracy.}
        \label{fig:res32_cluster8_converged_acc}
    \end{subfigure}
    \hfill
    \begin{subfigure}[t]{0.24\textwidth}
        \includegraphics[width=\textwidth]{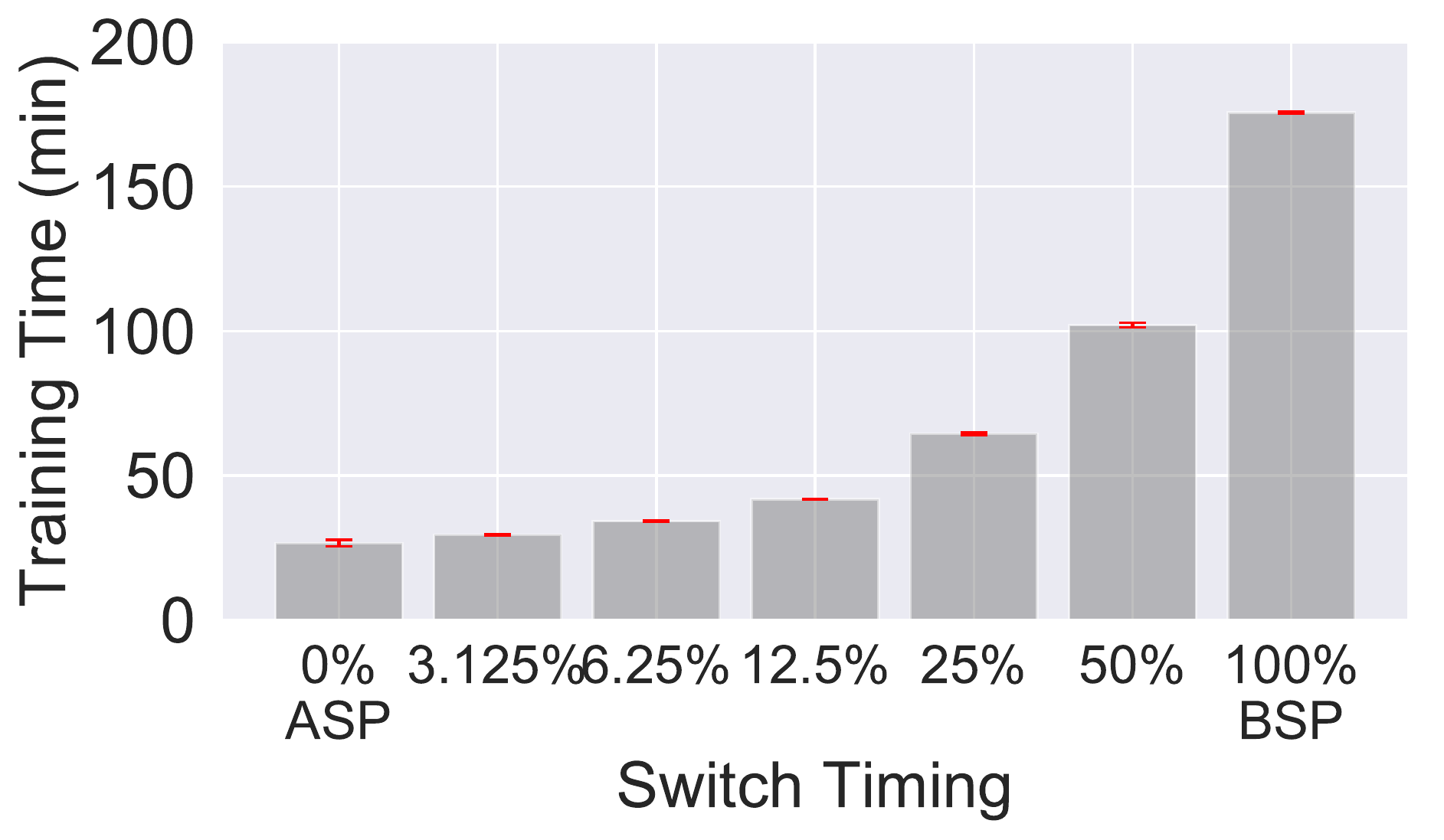}
        \caption{Total training time.}
        \label{fig:res32_cluster8_training_time}
    \end{subfigure}
    \caption{
    \textbf{Performance of exp. setup 1.} Using the policy of switching to ASP after completing 6.25\% workload with BSP, \sysname achieves similar converged accuracy and 80.5\% training time saving compared to training with BSP. 
    Between the range of 6.25\% to 50\%, switch timing has minimal impact on the converged accuracy but noticeable impact on training time.
    Dashed line marks the switch timing. 
    }
    \label{fig:res32_cluster8_eval}
\end{figure*}
\begin{figure*}[h!]
    \centering
    \begin{subfigure}[t]{0.24\textwidth}
        \includegraphics[width=\textwidth]{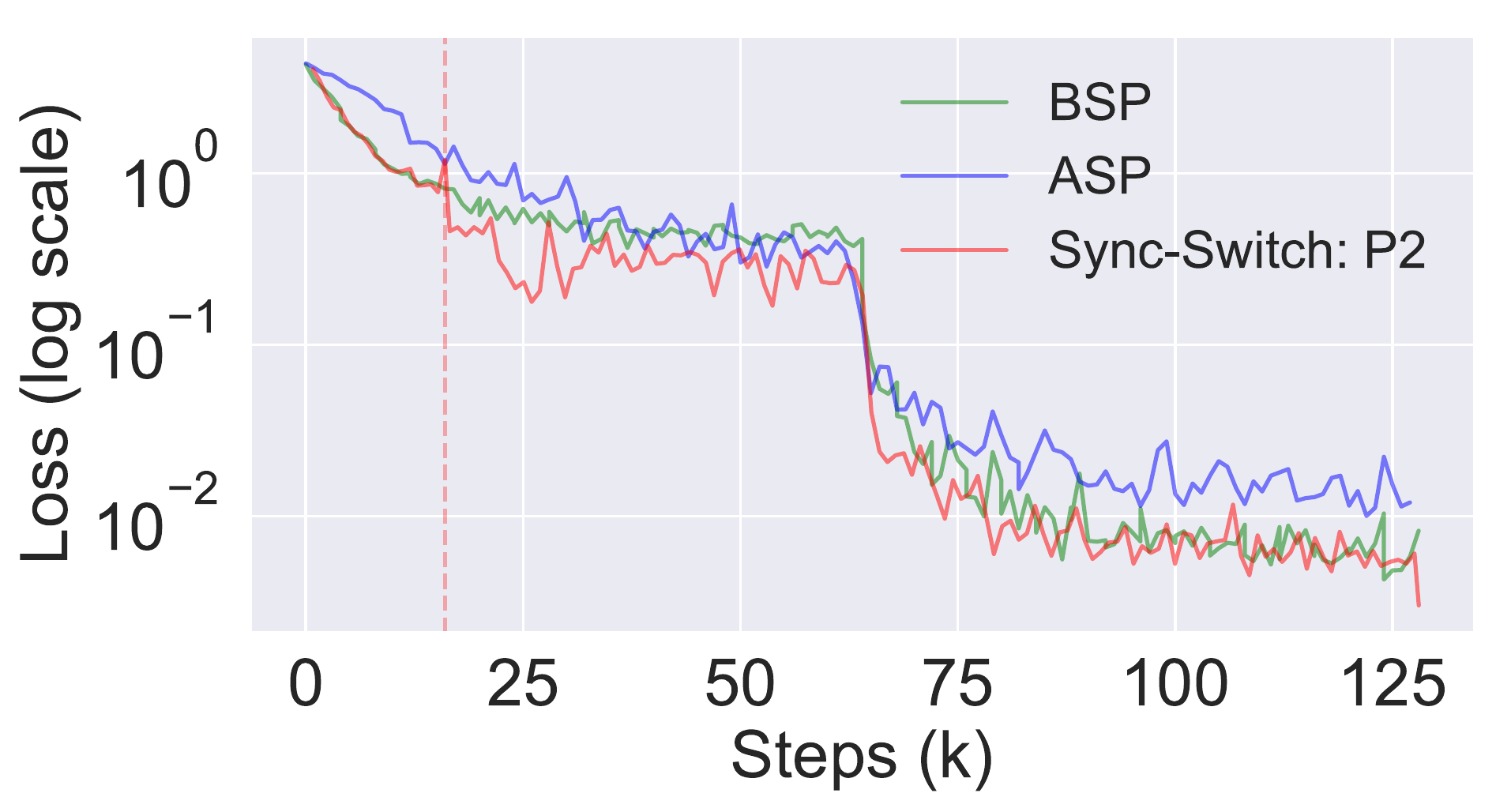}
        \caption{Training loss.}
        \label{fig:res50_loss_curve}
    \end{subfigure}
    \hfill
    \begin{subfigure}[t]{0.24\textwidth}
        \includegraphics[width=\textwidth]{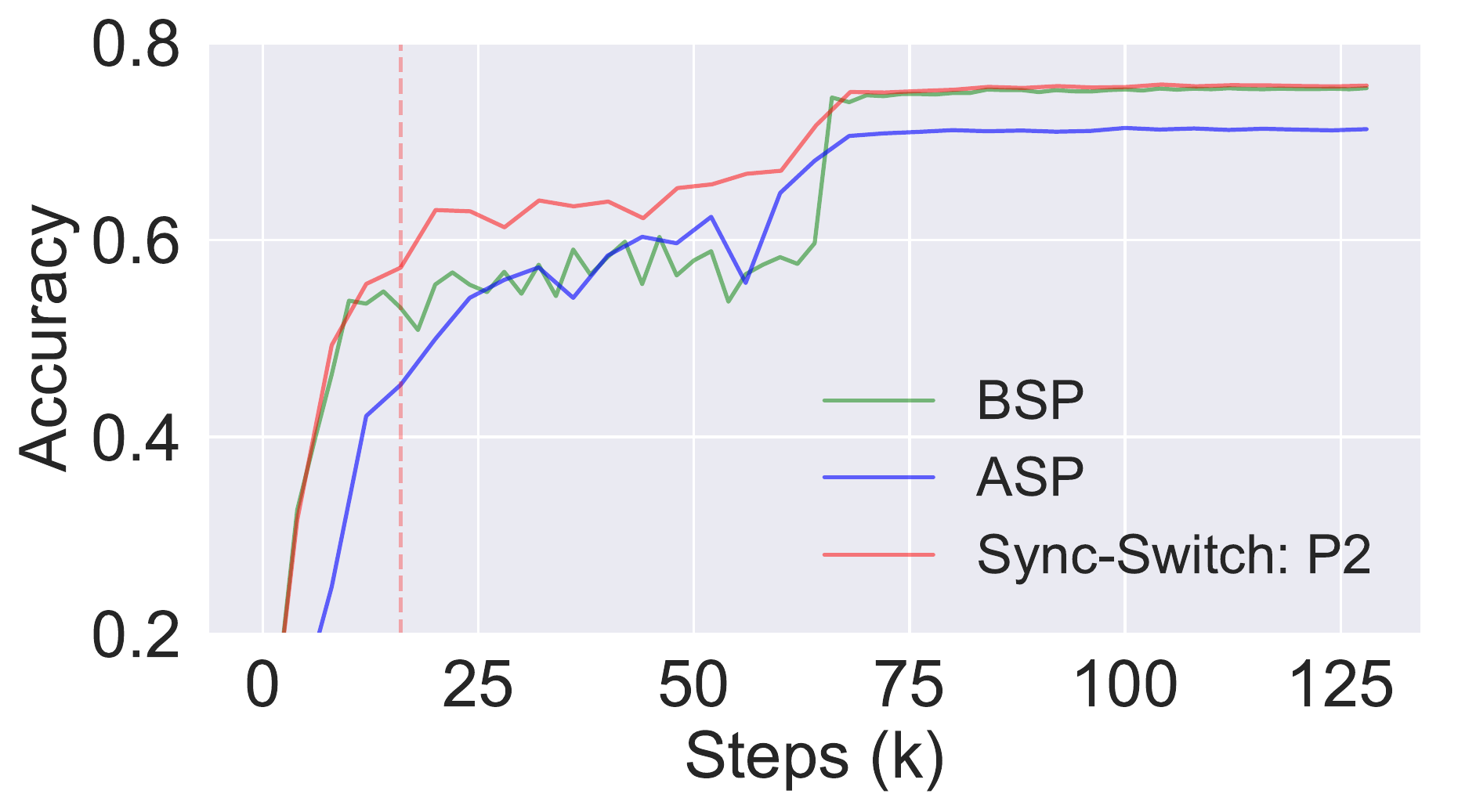}
        \caption{Test accuracy.}
        \label{fig:res50_acc_curve}
    \end{subfigure}
    \hfill
    \begin{subfigure}[t]{0.24\textwidth}
        \includegraphics[width=\textwidth]{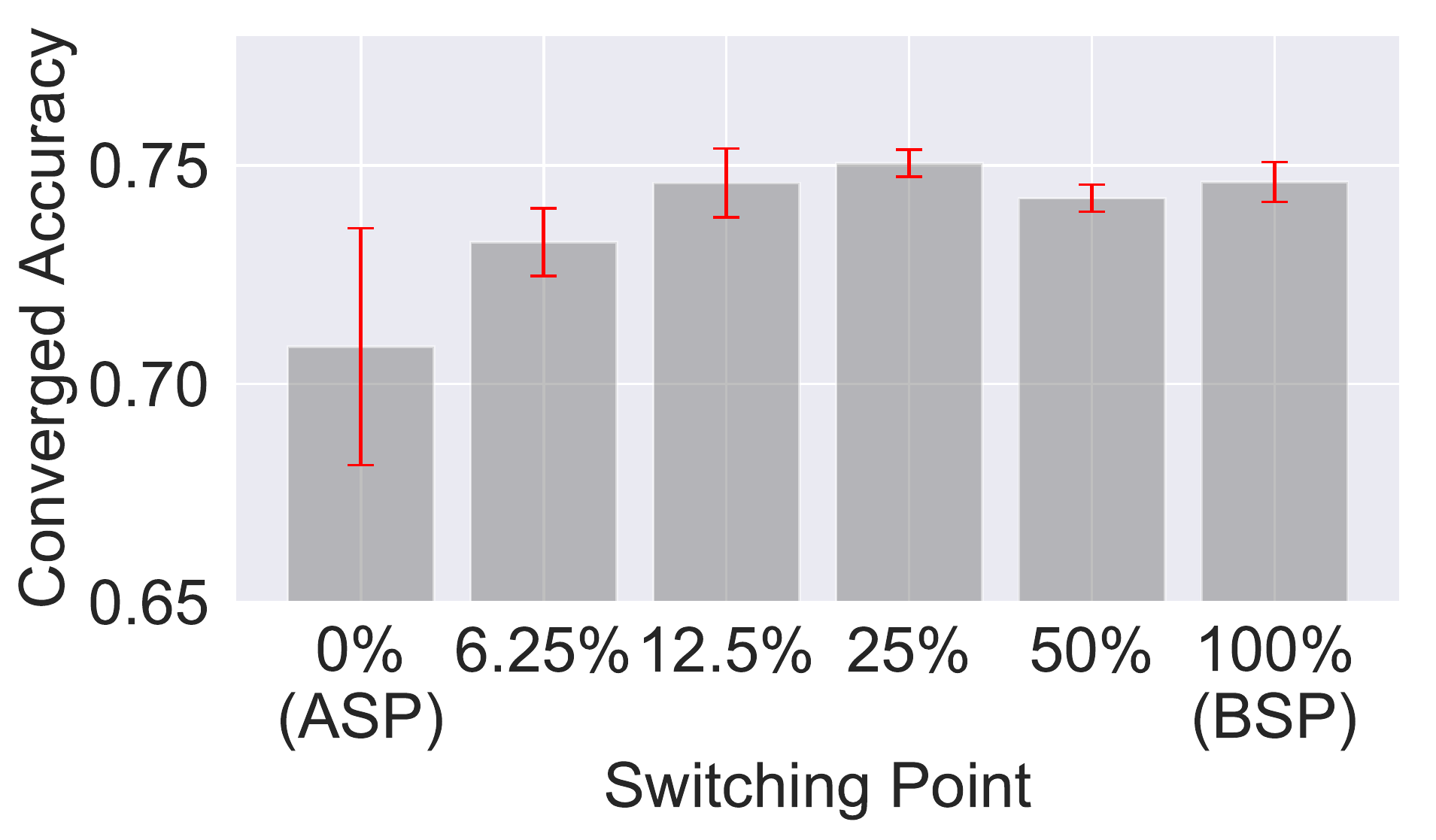}
        \caption{Converged accuracy.}
        \label{fig:res50_acc}
    \end{subfigure}
    \hfill
    \begin{subfigure}[t]{0.24\textwidth}
        \includegraphics[width=\textwidth]{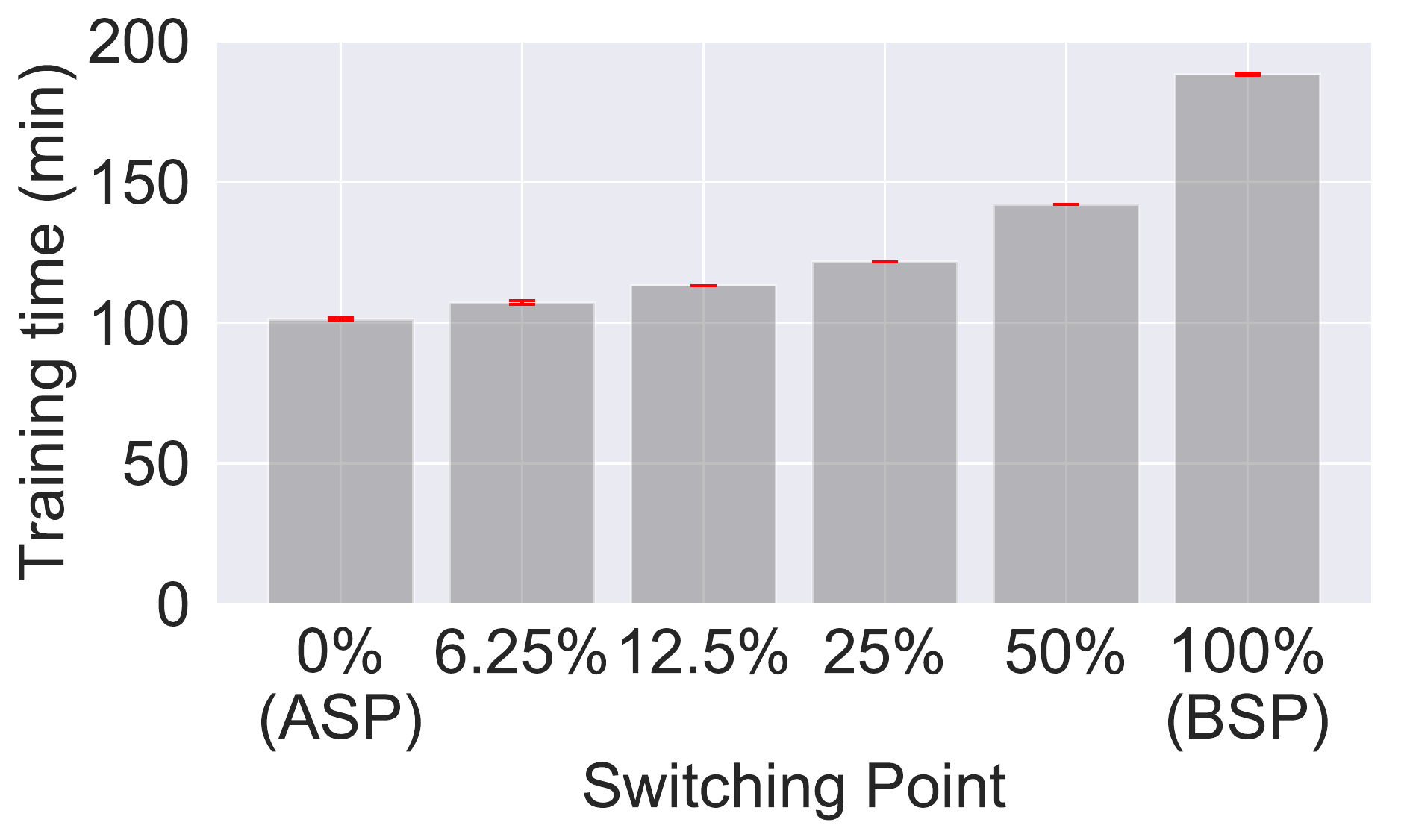}
        \caption{Total training time.}
        \label{fig:res50_time}
    \end{subfigure}
    \caption{
    \textbf{Performance of exp. setup 2.}
    With the policy of switching to ASP after completing 12.5\% steps using BSP, \sysname achieves similar converged accuracy with 39.9\% training time saving, compared to training with BSP.
    }
    \label{fig:res50_eval}
\end{figure*}
\begin{figure*}[t]
    \centering
    \begin{subfigure}[t]{0.24\textwidth}
        \includegraphics[width=\textwidth]{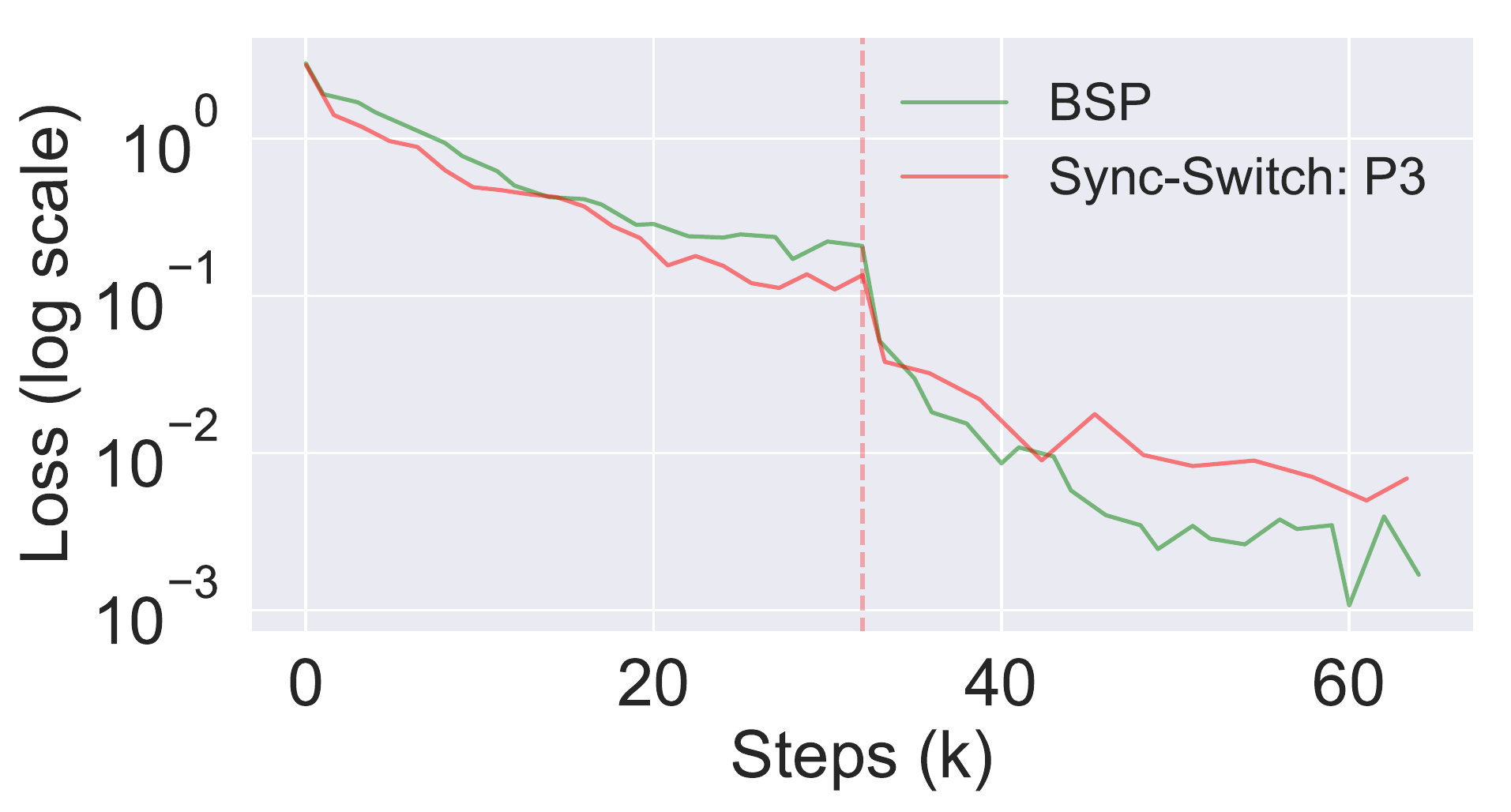}
        \caption{Training loss.}
        \label{fig:res32_cluster16_anytime_loss}
    \end{subfigure}
    \hfill
    \begin{subfigure}[t]{0.24\textwidth}
        \includegraphics[width=\textwidth]{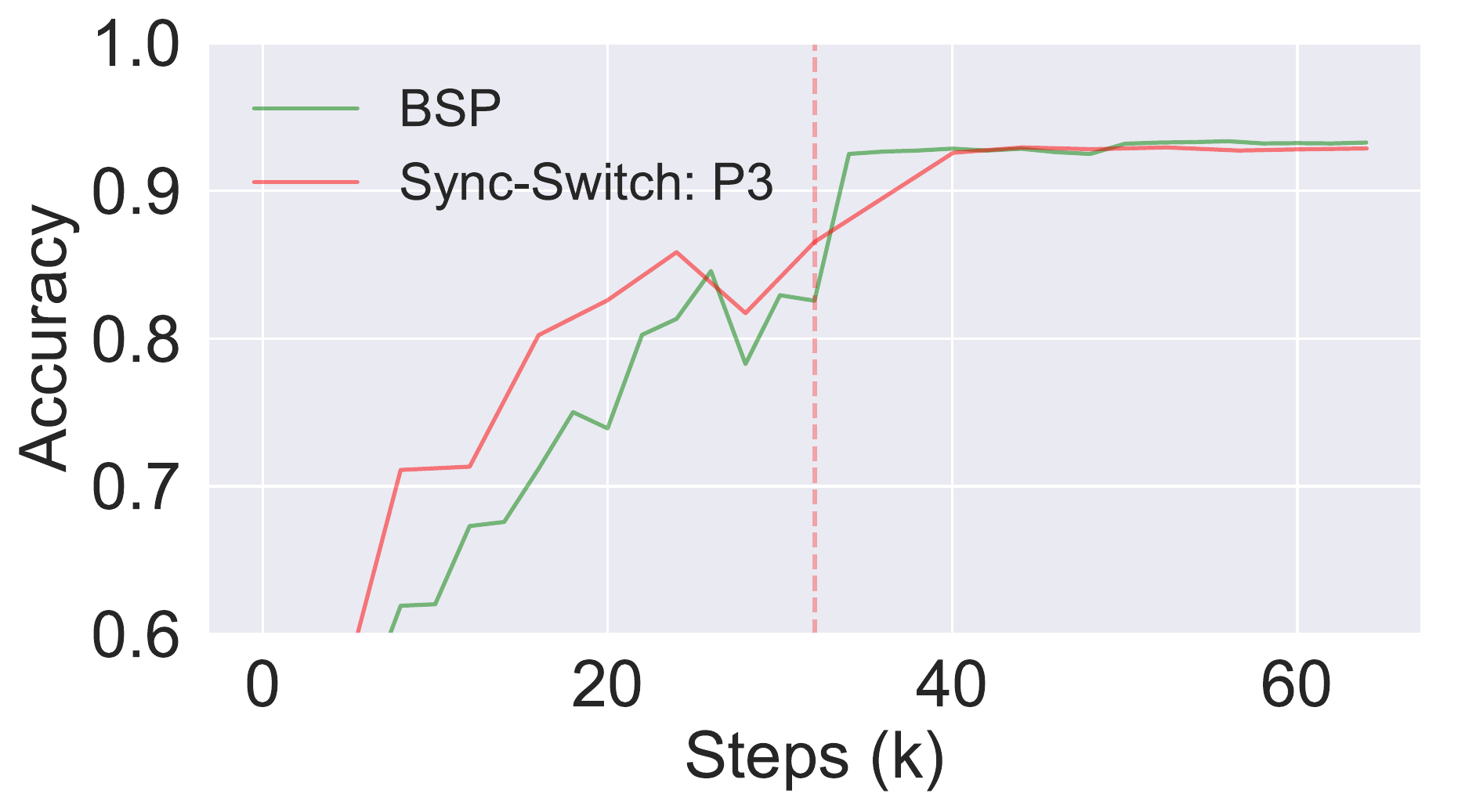}
        \caption{Test accuracy.}
        \label{fig:res32_cluster16_anytime_acc}
    \end{subfigure}
    \hfill
    \begin{subfigure}[t]{0.24\textwidth}
        \includegraphics[width=\textwidth]{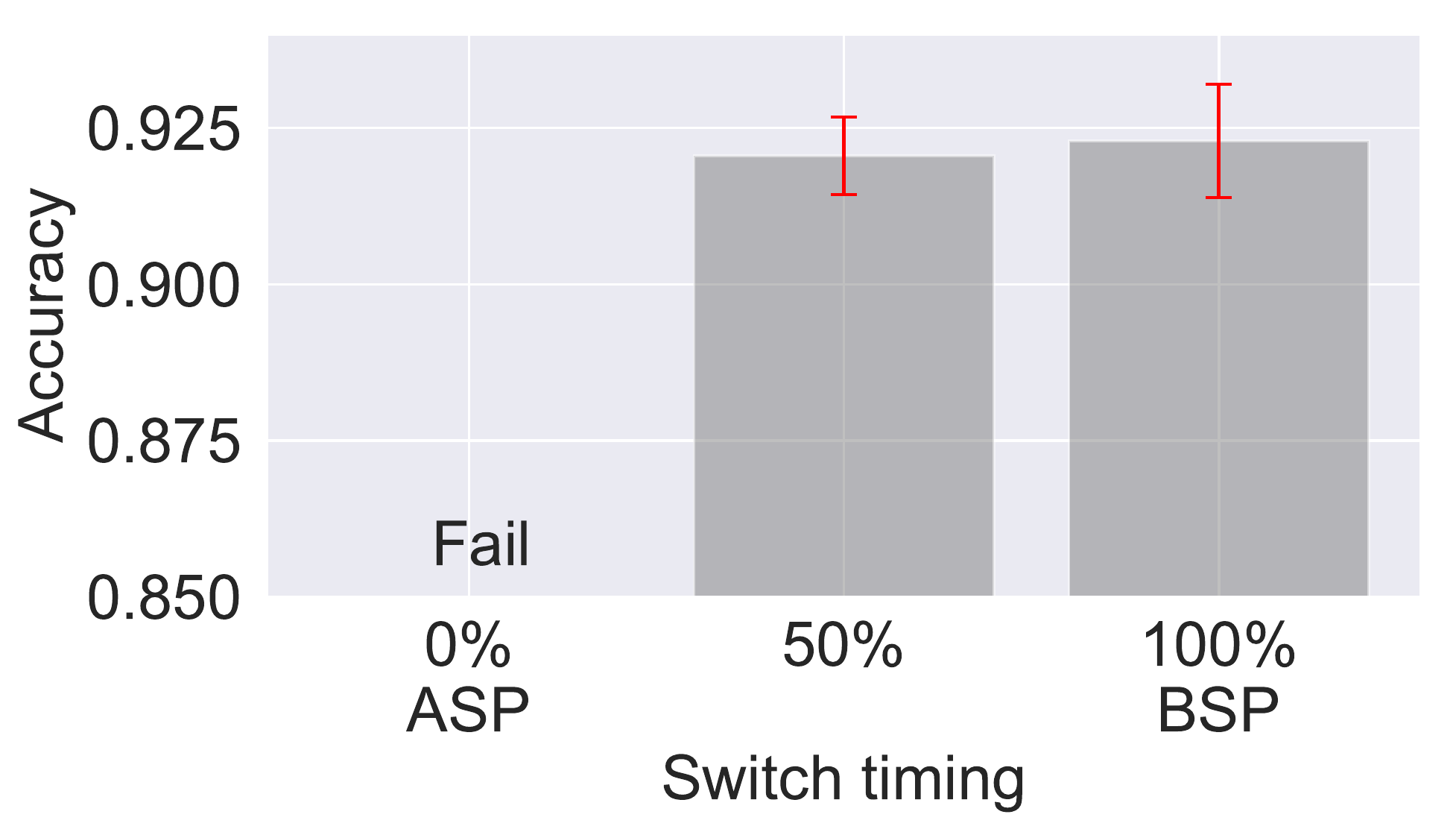}
        \caption{Converged accuracy.}
        \label{fig:res32_cluster16_test_acc}
    \end{subfigure}
    \hfill
    \begin{subfigure}[t]{0.24\textwidth}
        \includegraphics[width=\textwidth]{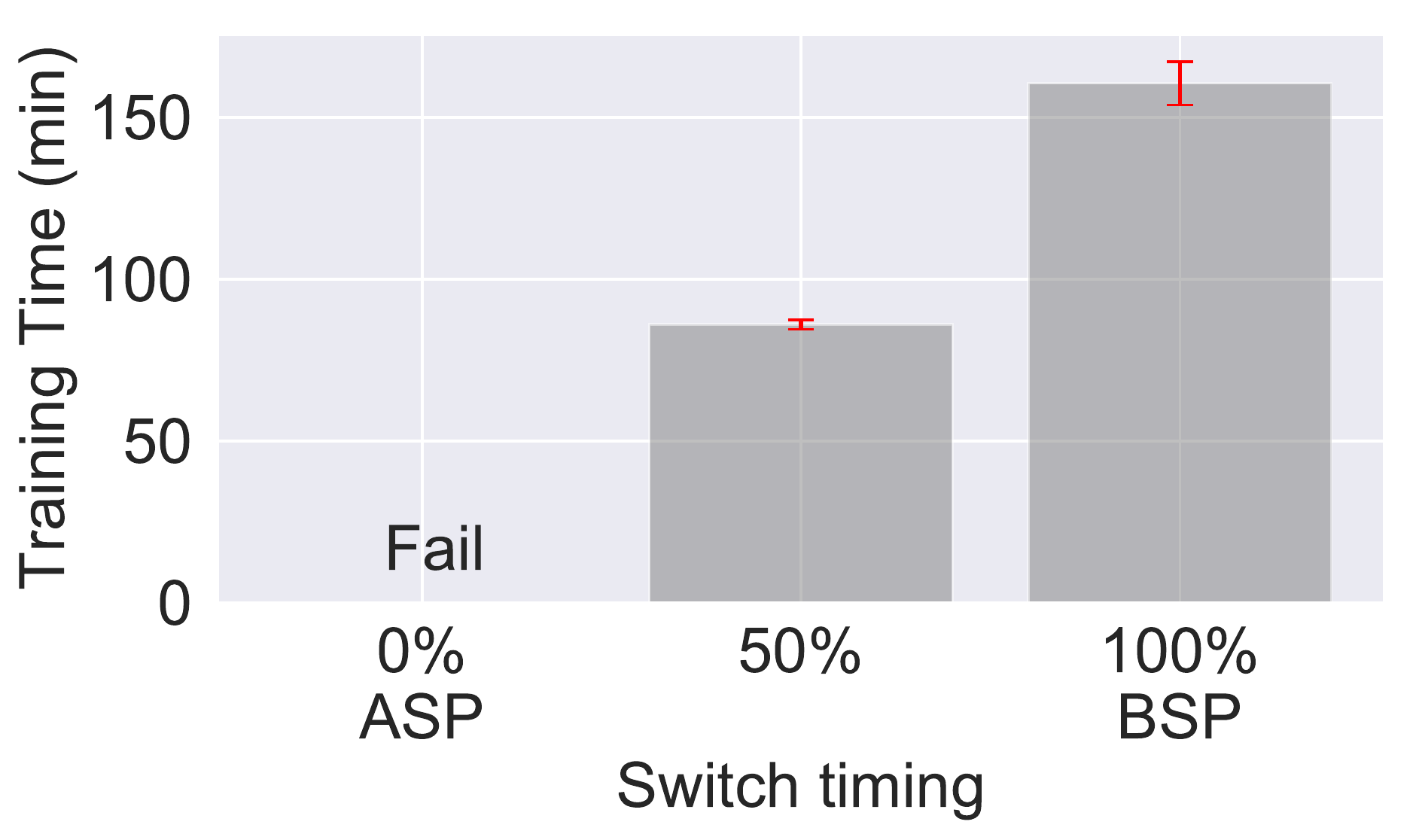}
        \caption{Total training time.}
        \label{fig:res32_cluster16_time}
    \end{subfigure}
    \caption{\textbf{Performance of exp. setup 3.}
    Due to the highly unstable convergence of stale gradients, ASP and switching before 50\% (first learning rate decay) diverged, leading to failed training.
    }
    \label{fig:res32_cluster16_eval}
\end{figure*}

\subsubsection{End-to-end Comparison}
Table~\ref{table:exp_summary_policies} and Figure~\ref{fig:eval_search_policy_overview} summarize the end-to-end training performance achieved by \sysname. In all setups evaluated, \sysname outperforms training exclusively with BSP and with ASP in total training time, TTA, and converged accuracy, respectively. 
For example, we observe that \sysname uses only 19.5\% of the training time while reaching similar converged test accuracy when compared to BSP;
additionally, \sysname improves converged accuracy by 2.5\% with only 1.28X the training time when compared to ASP (exp. setup 1).
The speedups achieved by \sysname are more prominent compared to 1.2-3.7X reported in prior work~\cite{jiang2019novel}; the high-quality converged accuracy is significant as recent innovations on models have similar levels of improvement, e.g., less than 2\%~\cite{gastaldi2017shake, chollet2017xception}.
The training speedup comes from \sysname only needs to train a small portion of workload using BSP and the competitive converged accuracy comes from identifying the most appropriate switch timing.

Furthermore, Figure~\ref{fig:res32_cluster8_eval} details the performance of training a ResNet32 on the CIFAR-10 dataset using an 8-GPU cluster. We plot training loss and test accuracy of the best runs for ASP, BSP, and \sysname in Figures~\ref{fig:res32_cluster8_anytime_loss} and \ref{fig:res32_cluster8_anytime_acc}.
Interestingly, training with BSP at the first 6.25\% steps allows \sysname to decrease the training loss faster and to lower values (the lower the better), compared to training exclusively with ASP. 
Additionally, even though \sysname does not decrease training loss to the same level as BSP, it still reaches the same converged accuracy. More importantly, \sysname only needs 25\% training time used by BSP to reach the same converged accuracy (i.e., a 4X TTA speedup).
Further, Figures~\ref{fig:res32_cluster8_converged_acc} and ~\ref{fig:res32_cluster8_training_time} draw the comparison to manual switching, i.e., switching at a static time point, and demonstrate \sysname's utility. 

Results of similar magnitude are also observed for the other two setups as shown in Figure~\ref{fig:res50_eval} and Figure~\ref{fig:res32_cluster16_eval}. Note that training ResNet32 with ASP (and training with BSP for less than 50\% steps) in a cluster of 16 GPU servers resulted in failed training due to the training loss divergence. The failed training sessions are likely caused by noisy gradients that are exacerbated with larger cluster sizes. This observation attests to an additional benefit provided by \sysname---being able to complete training in scenarios where ASP cannot.

%

 


\subsubsection{Generality Analysis of Our Observations}
%
%
\begin{figure}[t]
    \centering
    \begin{subfigure}[t]{0.24\textwidth}
        \includegraphics[width=\textwidth]{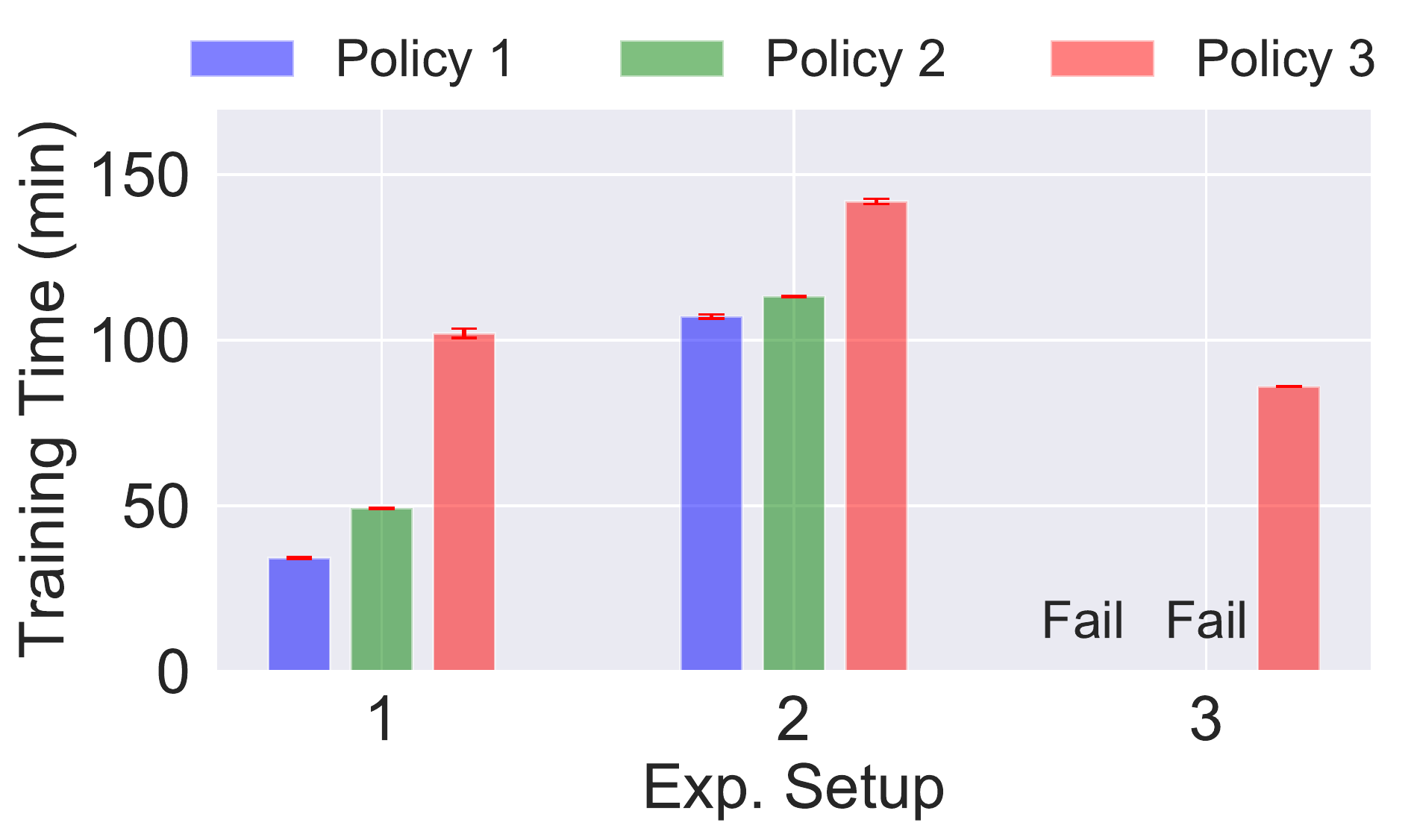}
        \caption{Total training time.}
        \label{fig:cross_comp_policy_time}
    \end{subfigure}
    \hfill
    \begin{subfigure}[t]{0.24\textwidth}
        \includegraphics[width=\textwidth]{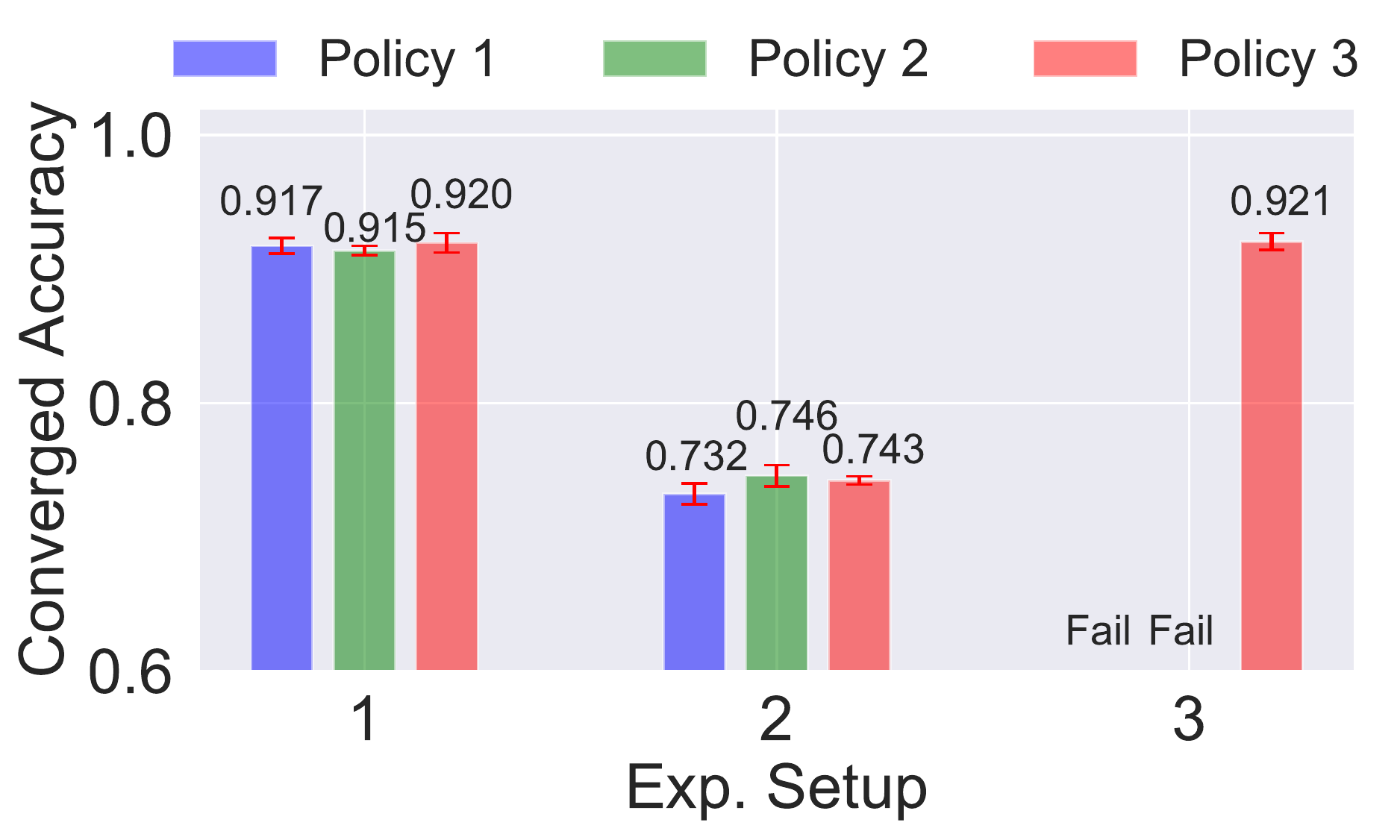}
        \caption{Converged test accuracy.}
        \label{fig:cross_comp_policy_time_accuracy}
    \end{subfigure}
    \caption{\textbf{Cross examination of \sysname policies in different experiment setups.}
    Policy $i$ represents the set of policies found by \sysname for experiment setup $i$ (as summarized in Table~\ref{table:exp_summary_policies}).
    }
    \label{fig:eval_cross_comp_policy}
\end{figure}
\begin{figure}[t]
    \centering
    \begin{subfigure}[t]{0.24\textwidth}
        \includegraphics[width=\textwidth]{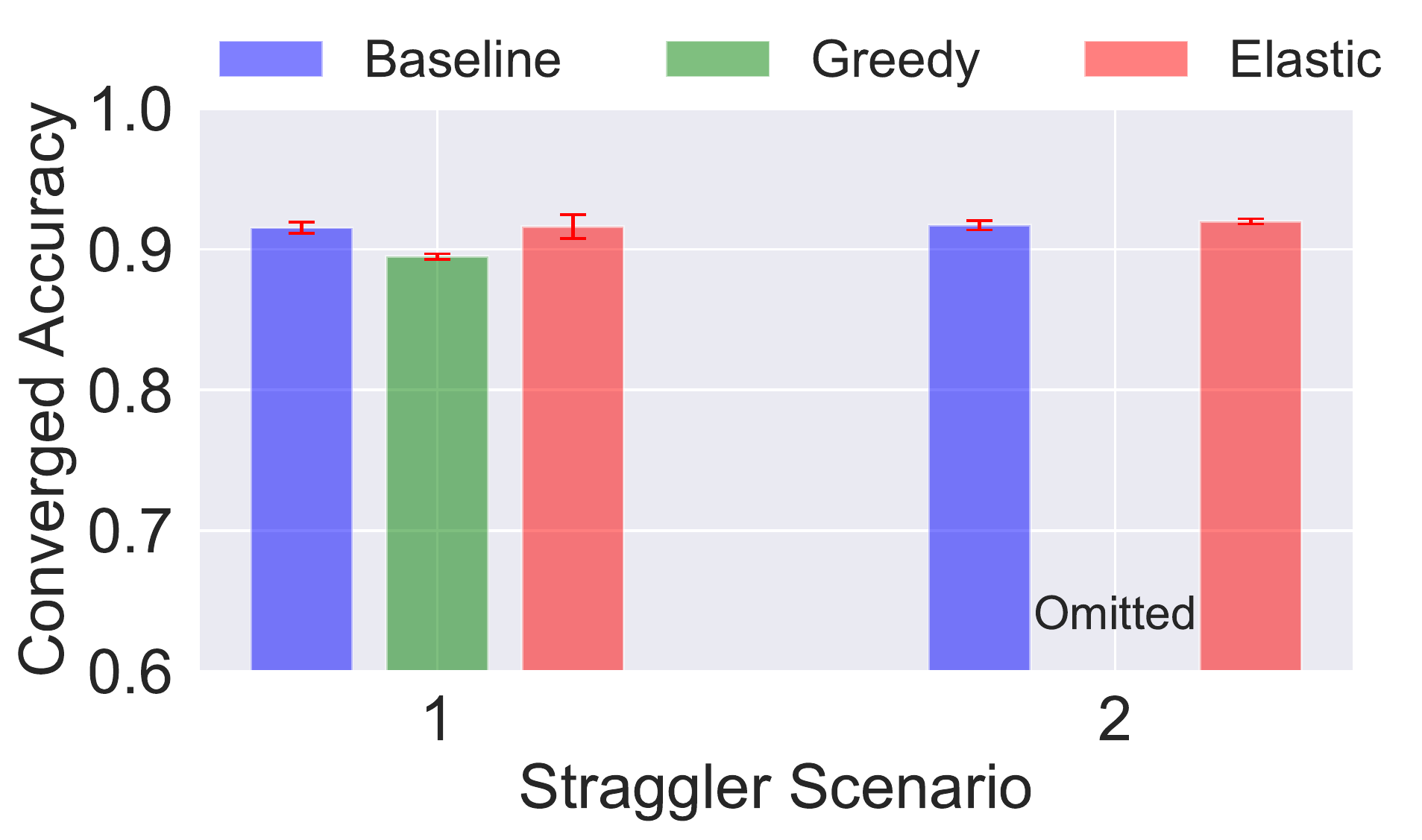}
        \caption{Converged accuracy.}
        \label{fig:eval_strag_acc}
    \end{subfigure}
    \hfill
    \begin{subfigure}[t]{0.24\textwidth}
        \includegraphics[width=\textwidth]{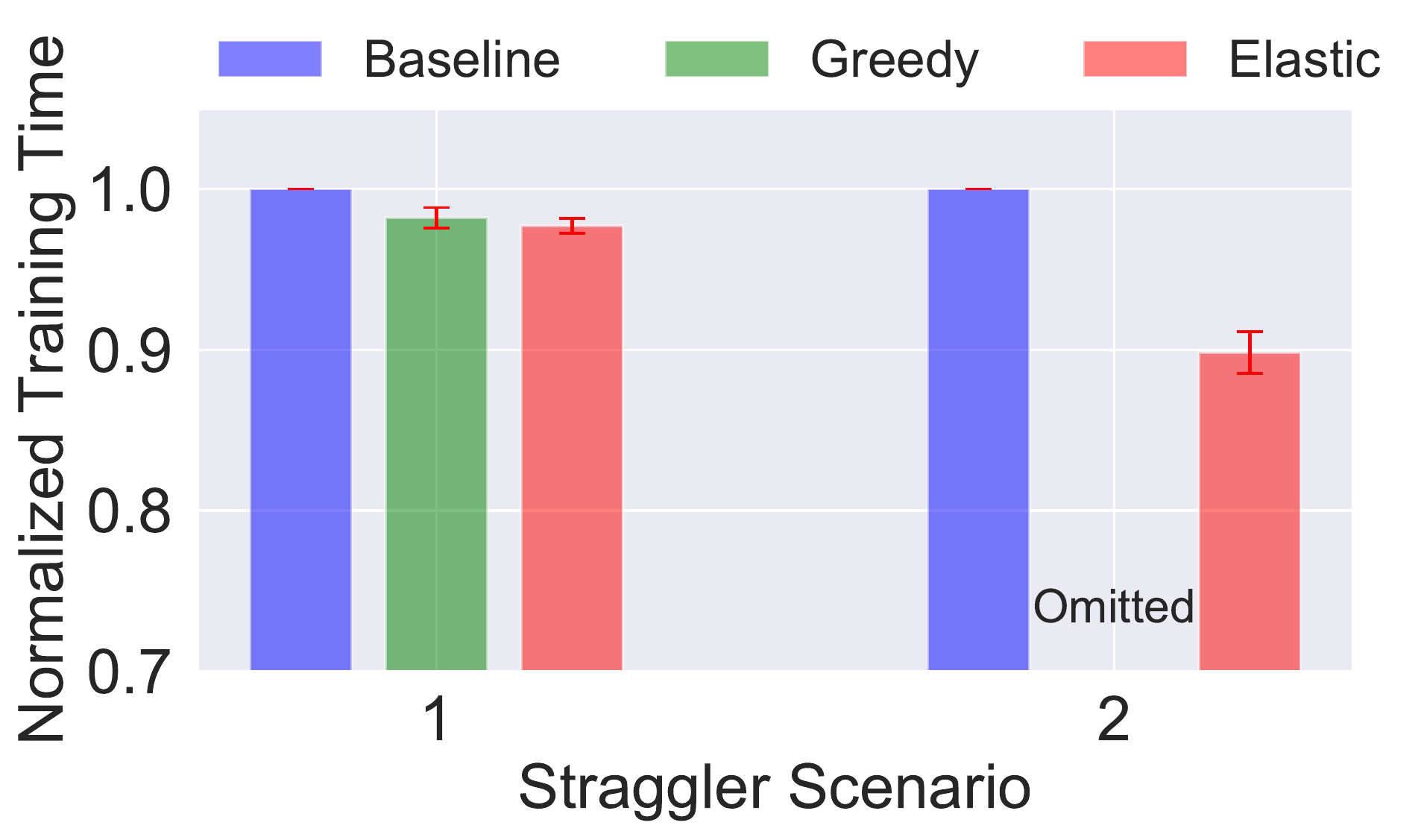}
        \caption{Normalized training time.}
        \label{fig:eval_strag_time}
    \end{subfigure}
    \caption{
    \textbf{Comparison of \sysname's straggler-aware policies.}
    We observe that the elastic-based policy preserves the converged test accuracy and has a 1.1X speedup compared to the baseline policy.
    }
    \label{fig:eval_strag}
\end{figure}

Figure~\ref{fig:eval_cross_comp_policy} compares the performance of directly using policies found for the other two setups to the third setup. We make the following key observations. 
First, for similar workloads (i.e., training ResNet32 on Cifar-10 vs. ResNet50 on Cifar-100), the cluster size seems to play an important role in determining the efficiency of the policies. Concretely, training with policy 2 in the experiment setup 1 achieves almost the same converge accuracy (91.7\% vs. 91.4\%) and uses about 33.0\% longer training time. The prolonged training time is expected as policy 2 requires more training to be done with BSP. In comparison, training with policy 3 under the same setup uses 3X of the training time with policy 1.  
Furthermore, by using the correct policy under experiment setup 3, the training successfully finished (without diverging) and achieved comparable test accuracy to using BSP while saving almost 46.4\% training time. As shown in Figure~\ref{fig:res32_cluster16_eval}, training with ASP failed without producing usable models. This observation further suggests the practical utility of \sysname.

\subsubsection{With Transient Stragglers}

We construct two transient straggler scenarios, mild and moderate, for experiment setup 1 to evaluate the effectiveness of the straggler-aware policies introduced in Section~\ref{subsubsec:online_policies}.
In the first scenario, we set the number of stragglers to be 1, the frequency of straggler occurrences to be 1, and inject the slowness by emulating the network latency to be 10ms. 
In the second scenario, we increase the number of stragglers, the frequency of straggler occurrences, and the degree of slowness to be 2, 4, and emulated with 30ms network latency. 

Figure~\ref{fig:eval_strag} compares the training performance with and without (baseline) applying our straggler-aware policies. We observe that when the straggler scenario is mild (scenario one), both straggler-aware policies adequately handle the potential performance degradation and even shorten the total training time by 2\%, compared to the straggler-agnostic baseline policy. Furthermore, we find that the greedy policy leads to a 2\% lower converge accuracy while policy two is able to maintain the high-quality converge accuracy. The accuracy degradation is most likely due to having to perform two extra switches, one to ASP and the other back to BSP, before the optimal timing.  
Based on our empirical observation, we conclude that the greedy policy does not work in conjunction with \sysname's existing baseline policies.   

In contrast, the elastic-based policy is proven to be effective even under moderate levels of slowness. In particular, we observe that this policy not only achieves a similar level of converged accuracy but also leads to a 1.11X speedup. This further suggests that the better course of action is to train without the transient stragglers than to block the remaining cluster nodes from making progress when training BSP.

\subsection{Overhead of \sysname}


\subsubsection{Binary Search Cost} 
\label{subsec:search_overhead}

\begin{table}[t]
\centering 
\footnotesize
\begin{tabular}{l|rrrr}
\toprule
\textbf{Search Setting} &
  \textbf{Cost} &
  \textbf{Amortization} &
  \multicolumn{1}{c}{\begin{tabular}[c]{@{}r@{}}\textbf{Effective}\\ (vs. BSP) \end{tabular}} &
  \begin{tabular}[c]{@{}r@{}}\textbf{Success} \\ \textbf{Probability}\end{tabular} \\
\midrule 
\color{gray}{(Exp.1, No, 5, 5)}     & 12.71X & 15.79 & 1.97X       & 100\%        \\
  (Exp.1, No, 3, 3)   & 7.62X & 9.47 & 1.97X            & 99.2\%       \\
  (Exp.1, Yes, 0, 3)        & 4.63X & 5.75 & 2.59X            & 100\%        \\
\midrule
\color{gray}{(Exp.2, No, 5, 5)}      & 17.86X & 44.81 & 1.12X            & 100\%        \\
(Exp.2, No, 4, 4) & 14.28X & 35.83 & 1.12X            & 93.4\%       \\
(Exp.2, Yes, 0, 4)          & 9.05X & 22.71 & 1.17X            & 100\%        \\
\midrule
\color{gray}{(Exp.3, No, 5, 5)}       & 7.68X & 16.54 & 1.30X            & 100\%        \\
(Exp.3, No, 3, 3)   & 4.61X & 9.93 & 1.30X            & 100\%        \\
(Exp.3, Yes, 0, 1)          & 0.54X & 1.16 & 1.87X            & 100\%        \\
\bottomrule
\end{tabular}
\caption{
\textbf{Binary search cost analysis.}
We define a search setting as job recurrence, number of BSP trainings, and number of candidate policy trainings.
%
}
\label{table:search_cost_res32}
\vspace{-1em}
\end{table}



To quantify the overhead of our binary search-based algorithm (in search time) in different training scenarios, i.e., with recurring jobs and fewer measurement runs, we use all our training logs and simulate each search setting 1000 times with the accuracy threshold of 0.01.
%
Table~\ref{table:search_cost_res32} details the cost-performance trade-off (more results are available in the appendix). 
If the job is recurring, the search cost can be reduced by up to 5X the BSP training.
However, when facing a new training job, it is best to at least repeat the BSP runs 3 times. We further observe that with  too few BSP training, the search setting often ends up with significantly lower success probability (e.g., 56.8\% to 82.3\%) in finding the same switching timing as the baseline setting. 
%

Additionally, we analyze the amortized cost, measured by the number of job recurrences, and the effective training ratio, measured by the multiples of BSP training sessions. The former provides further justification of the cost-effectiveness of our binary search algorithm and the latter signifies the potential information gains. As an example of the search setting (No, 3, 3) in Table~\ref{table:search_cost_res32}, if a job needs to be trained for more than 9 times, a very likely event given the trial-and-error nature of deep learning training, the corresponding search cost is then amortized. Moreover, the search process itself also produces almost 2X valid training sessions compared to training with BSP. In summary, our analysis shows that the search cost is reasonably low and can be further reduced by continuously using \sysname.

\subsubsection{Runtime Overhead}
\label{subsec:practical_overhead}

\begin{table}[t]
\centering
\footnotesize
\begin{tabular}{r|rrrr}
\toprule
\textbf{Cluster} & \textbf{Actuator Exec.} & \textbf{Init. (s)} & \textbf{Switching (s)} & \textbf{Total (s)} \\
\midrule
\multirow{2}{*}{8 K80} & Sequential & 157  & 90 & 247 \\ 
                    & Parallel (Ours) & 90   & 36  & 126 \\
\multirow{2}{*}{16 K80}   &  Sequential & 268 & 165  & 433 \\ 
                      & Parallel (Ours)  & 128  & 53 & 181  \\
\bottomrule
\end{tabular}
\caption{
\textbf{\sysname Overhead.} We measure both the initialization time, i.e., the time taken to setup the training cluster, and the switching overhead, when training ResNet32 using \sysname.
%
%
}
\label{table:framework_overhead}
\vspace{-2em}
\end{table}



We quantify the overhead of using \sysname to perform distributed training. Our measurements are based on training ResNet32 on the CIFAR-10 dataset, as the training  framework overhead is largely workload-independent~\cite{Li2020-bk}. Table~\ref{table:framework_overhead} shows the total time spent by \sysname for  initializing the cluster and switching to a different synchronization protocol. 
First, initializing a cluster of twice the workers takes 1.42X the time of initializing a cluster of 8 workers. 
Note that one can expect to have similar initialization time even with just the vanilla TensorFlow~\cite{Li2020-bk}. 
Second, by having a configuration actuator that propagates distributed training tasks in parallel, \sysname reduces both the initialization time and switching overhead by 2X and 3.1X, respectively. 
Third, the switching overhead can be as low as 36 seconds, about 1.7\% of the time taken to train the model with \sysname. 
In summary, \sysname incurs low switching overhead that increases sub-linearly with the cluster size.  
\section{Related Work}
\label{sec:related}


\para{Distributed Synchronization Protocols.} 
Researchers have designed many synchronization protocols that can be roughly categorized as BSP~\cite{Gerbessiotis1994-wy}, ASP~\cite{dean2012large}, and semi-synchronous protocols~\cite{ho2013more, zhao2019dynamic}, that trade-off the training throughput and accuracy of distributed DL training.
These studies all focus on improving the synchronization protocols for distributed SGD, by exposing mechanisms and policies to control the model staleness. In contrast, our work focuses on determining the best way to utilize existing protocols and can be used in conjunction with new synchronization protocols.
Compared to semi-synchronous protocols such as SSP and DSSP, our work leads to good converged accuracy and does not require users to tune extra hyper-parameters. 
In addition to directly modify the synchronization protocols, researchers also look at using different synchronizations for different cluster nodes to account for the heterogeneous performance caused by network and GPU servers~\cite{jiang2019novel,hsieh2017gaia}. For example, Gaia used synchronous training for nodes running inside the same datacenter and fully asynchronous training for inter-datacenter communication~\cite{hsieh2017gaia}. 
Additionally, Dutta et al.~\cite{dutta2020slow} 
introduced a number of SGD variants where the synchronization degree is controlled by a new hyper-parameter.
Our work is similar in that we will also use different synchronizations for a given training session but with the key difference of deriving the policies for hybrid synchronization.  


\para{Network Optimization for Distributed Training.} The iterative process of deep learning makes network an important bottleneck as not only the model parameters but also the gradients need to be transferred periodically. To combat the impact on training performance without impacting the model quality, prior work explored various techniques that aim to reduce the communication costs via gradient sparsification or compression. For example, by only sending the large gradients, Aji et al. used a heuristic sparsification scheme and showed a speed gain of 22\%~\cite{Aji2017-yr}. 
Terngrad and QSGD improved the network communication efficiency by reducing the gradients to a few numerical levels~\cite{Wen2017-uf,Alistarh2017-ss}. These efforts are orthogonal to our work but might be combined with \sysname to achieve further training speedup. 

\section{Conclusion}
\label{sec:conclusion}

Using the right distributed synchronization protocol at the right time can significantly improve the training throughput and produce models with good test accuracy. In this paper, we devised the first set of adaptive policies, including offline and online ones, that make such decisions and evaluated their effectiveness in a prototype system called \sysname in Google Cloud.
We found that training with BSP for only a small portion of time and then switching to ASP delivers models of comparable converged accuracy using much shorter time, compared to training with BSP.
%
%
%
For recurring training jobs, a prevalent scenario in deep learning due to its trial-and-error nature, we used an offline approach to find the optimal switch timing.
%
To ensure that switching from BSP to ASP does not lead to undesirable side effects, we additionally specified a configuration policy that describes how to adjust critical hyper-parameters. 
We showed that \sysname improved the total training time and the time-to-accuracy by up to 5X and 4X while achieving similar test accuracy through real-world experiments, compared to training exclusively with BSP. 
Further, with the elastic-based policy, \sysname can effectively circumvent the performance degradation caused by transient stragglers and instead leads to a 1.1X speedup under moderate slowdown scenarios.
Additionally, the benefits brought by \sysname come with reasonably low overhead, e.g., a search overhead that can be amortized with jobs recurring a few times and a switching overhead in the order of tens seconds. 
%



\section{Acknowledgements}
We would like to thank all anonymous reviewers for their insightful comments. This work is supported in part by National Science Foundation grants \#1755659 and \#1815619, National Natural Science
Foundation of China (61802377) and Youth Innovation
Promotion Association at CAS, and Google Cloud Platform Research credits, 



\balance
\bibliographystyle{IEEEtran}
\bibliography{bib/arxiv_version}

\appendices
\section{Additional remarks for theoretical explanations}
\label{sec_loss_landscape}

This section provides additional remarks for \sysname protocol policy design described in Section~\ref{subsec:ordering}.

\begin{remark}[Population loss landscape at different scales]\label{rem_population_loss}
 There is much empirical and heuristic evidence that the  population loss is in fact much smoother than the training loss. 
 For instance, it has been observed that sharp local minima of the training loss -- minima where the training loss  is only low in a small region near the minimum point -- oftentimes do not correspond to a low test loss \cite{hochreiter1995simplifying, hochreiter1997flat, keskar2016large, smith2018bayesian, chaudhari2019entropy}.
On the other hand, 'flat' local minima of the training loss-- local minima where the training loss remains low in a wide region containing the local  minimum point-- in many cases also minimize the test loss.
Moreover, it has recently been shown that minimizing a 'smoothed' version
of the training loss can lead to a test error that is as low as, or oftentimes
even lower than, if one were to minimize the training loss itself. 
Here the smoothed loss function is the convolution of the loss function with
the distribution of the batch gradient noise (the variance of the noise depends
on the batch size) \cite{kleinberg2018alternative}.

This evidence suggests that the population loss is much smoother than the training loss, and is in fact closer to a 'smoothed' version of the training loss convolved with a the distribution of the batch gradient noise rather than the original training loss. This suggest that our population loss will have a landscape that is smooth at a finer scale on the order of the learning rate used later in the training.  On the other hand, at coarser scales--on the order of the large learning rate used earlier in the training--we expect the landscape of the population loss to be much closer to the training loss. In other words, earlier in the training, when a large step size is used, the change in the training loss and population loss at each step should be roughly the same, and the ratio of the change in the population loss to the change in training loss should be close to 1. Later in the training, when a much smaller learning rate is used, this ratio could be very far from 1.
\end{remark}

\begin{remark}[Can switching from BSP to ASP also minimize the training loss?]\label{rem_can_BSP_to_ASP_minimize_training_loss}
In the previous discussion we have used the fact that the population loss is oftentimes much smoother than the training loss.
For this reason, even if one only uses stale gradients later in the training when the algorithm takes smaller step sizes, the gradient of the training loss may still change rapidly at each step of the algorithm despite the fact that the gradient of the population loss may not be changing as quickly.
This suggests that, while using ASP later in the training may allow one to effectively minimize the population loss, it may still not allow one to minimize the training loss.
This is exactly what we observe in our experiments: while starting with BSP and then switching to ASP allows one to minimize the test error (and hence the population loss) as effectively as using static BSP, the training loss remains much higher than in experiments where static BSP is used.

\end{remark} 

\begin{remark}[Why not start with ASP and then switch to BSP?] \label{rem_why_not_ASP_to_BSP}
As noted in the previous discussion, using ASP early in the training may cause the algorithm to be unstable, preventing ASP from effectively decreasing the loss value.
Thus, even if one starts with ASP and switches to BSP, the time spent running ASP early in the training is effectively wasted time.
This means that, starting with ASP and then switching to BSP does not allow for any speedup over static BSP (and may even lead to a slightly slower training time than static BSP due to the time wasted running ASP at the beginning of the training).

Moreover, in our experiments we observe that starting with ASP and switching to BSP can cause the  loss value to get ``stuck" at a relatively high value for a very long time.
This is likely because, if one starts with ASP and then switches to BSP after decaying the learning rate, the algorithm may still be in a region with a high loss value at the epoch when the learning rate is decreased.
 Since a lower learning rate can cause SGD to take a much longer time to escape saddle points of the loss function \cite{smith2017cyclical}, starting the training with ASP can cause the algorithm to get stuck for a very long time near a saddle point with a high loss value even if one then switches to BSP (Figure \ref{fig:cartoon_ASP_to_BSP}), .
\end{remark}

\section{Pseudo Code for Our Binary Search Algorithm}
\label{appendix:sec:pseudo}

\begin{algorithm}[h]
\caption{Our Binary Search-based Algorithm for Deriving Timing Policies}
\begin{algorithmic}[1]
\STATE \textbf{Inputs:} Accuracy threshold $\beta$, num. of settings $M$, runs per setting $R$, target accuracy $A$ (optional)
\IF{$A$ is not provided}
\STATE Train the model with BSP $R$ times and record converged accuracy: $\alpha_1 \cdots \alpha_{r}$
\STATE Set $A = \frac{1}{R}\sum_{i=1}^{R}\alpha_r$
\ENDIF
\STATE $upper = 100, lower = 0, m = 0, \alpha' = 0$
\WHILE{$m < M$}
\STATE $r = 0$
\STATE $switching\_timing = \frac{(upper + lower)}{2}$
\WHILE{$r < R$}
\STATE Train $switching\_timing\%$ of workload with BSP, then switch to ASP 
\STATE Record the converged accuracy $\alpha_r$
\STATE $\alpha' = \alpha' + \alpha_r$
\STATE $r = r + 1$
\ENDWHILE
\IF{$\frac{\alpha'}{R} \in [A - \beta, A + \beta ]$}
\STATE $upper = switching\_timing$
\ELSE
\STATE $lower = switching\_timing$
\ENDIF
\STATE $m = m + 1$
\ENDWHILE
\end{algorithmic}
\label{alg:binary}
\end{algorithm}


In Algorithm~\ref{alg:binary}, we specify a limit for number of settings $M$ we want to explore, since from the observation in our empirical experiment, the speedup provided for switching earlier over a small percentage of workload is negligible. $\beta$ is the margin of error for converged accuracy due to the stochastic nature of deep learning and distributed training, and should either be specified by the user or set automatically (see Section~\ref{sec:eval} for details). To further reduce variance, we set the number of runs for each switch point $R$. A large $R$ can reduce sub-optimal results, but increase the search cost. In reverse, a small $R$ can reduce search cost, but increase the possibility of sub-optimal results. $A$ is the target converged accuracy in the constraints.

\section{Additional Results for Binary Search-based Overhead Analysis}
\label{subsec:appendix:sim_results}

Table~\ref{table:search_cost_res32_detail}, Table~\ref{table:search_cost_res50}, and Table~\ref{table:search_cost_cluster16} provide the complete simulation results to what was presented in Section~\ref{subsec:search_overhead}.

\begin{table*}[h!]
\centering 
\footnotesize
\begin{tabular}{r|rrrr}
\toprule
\multicolumn{1}{c|}{\begin{tabular}[c]{@{}r@{}}\textbf{Search Setting}\\ (Recurring, BSP runs, candidate runs)\end{tabular}} &
  \textbf{Search Cost} &
  \multicolumn{1}{c}{\begin{tabular}[c]{@{}r@{}}\textbf{Amortized}\\ (\# of recurrence)\end{tabular}} &
  \multicolumn{1}{c}{\begin{tabular}[c]{@{}r@{}}\textbf{Effective Training}\\ (vs. BSP) \end{tabular}} &
  \begin{tabular}[c]{@{}r@{}}\textbf{Success} \\ \textbf{Probability}\end{tabular} \\
\midrule 
Baseline: (No, 5, 5)     & 12.71X & 15.79 & 1.97X       & 100\%        \\
(No, 4, 4)  & 10.17X & 12.63 & 1.97X            & 100\%        \\
(No, 3, 3)    & 7.62X & 9.47 & 1.97X            & 99.2\%       \\
(No, 2, 2)  & 5.07X & 6.30 & 1.97X            & 82.3\%       \\
(No, 1, 1)   & 2.48X & 3.08 & 2.02X            & 56.8\%       \\
(No, 1, 5)  & 8.71X & 10.82 & 2.41X            & 80.4\%       \\
(No, 1, 4)  & 7.16X & 8.90 & 2.37X            & 78.7\%       \\
(No, 1, 3)  & 5.61X & 6.97 & 2.32X            & 78\%         \\
(No, 1, 2)  & 4.06X & 5.04 & 2.22X            & 69.4\%       \\
(Yes, 0, 5)         & 7.71X & 9.58 & 2.59X            & 100\%        \\
(Yes, 0, 4)      & 6.17X &  7.67 & 2.59X            & 100\%        \\
(Yes, 0, 3)        & 4.63X & 5.75 & 2.59X            & 100\%        \\
(Yes, 0, 2)           & 3.07X & 3.81 & 2.61X            & 79\%         \\
(Yes, 0, 1)           & 1.50X & 1.86 & 2.67X            & 56.6\%       \\ 
\bottomrule
\end{tabular}
\caption{
\textbf{Cost and performance analysis for experiment setup one.}
}
\label{table:search_cost_res32_detail}
\end{table*}

\begin{table*}[h!]
\centering 
\footnotesize
\begin{tabular}{r|rrrr}
\toprule
\multicolumn{1}{c|}{\begin{tabular}[c]{@{}r@{}}\textbf{Search Setting}\\ (Recurring, BSP runs, candidate runs)\end{tabular}} &
  \textbf{Search Cost} &
  \multicolumn{1}{c}{\begin{tabular}[c]{@{}r@{}}\textbf{Amortized}\\ (\# of recurrence)\end{tabular}} &
  \multicolumn{1}{c}{\begin{tabular}[c]{@{}r@{}}\textbf{Effective Training}\\ (vs. BSP) \end{tabular}} &
  \begin{tabular}[c]{@{}r@{}}\textbf{Success} \\ \textbf{Probability}\end{tabular} \\
\midrule 
Baseline: (No, 5, 5)     & 17.86X & 44.81 & 1.12X            & 100\%        \\
(No, 4, 4) & 14.28X & 35.83 & 1.12X            & 93.4\%       \\
(No, 3, 3)   & 10.71X & 26.87 & 1.12X            & 85.4\%       \\
(No, 2, 2)   & 6.98X & 17.51 & 1.15X            & 67.3\%       \\
(No, 1, 1)   & 3.26X & 8.18 & 1.23X            & 37.3\%       \\
(No, 1, 5)   & 12.12X & 30.41 & 1.41X            & 79.8\%       \\
(No, 1, 4)   & 10.04X & 25.19 & 1.29X            & 59.2\%       \\
(No, 1, 3)   & 7.75X & 19.45 & 1.29X            & 49\%         \\
(No, 1, 2)   & 5.65X & 14.18 & 1.24X            & 48.2\%       \\
(Yes, 0, 5)         & 11.10X & 27.85 & 1.17X            & 100\%        \\
(Yes, 0, 4)          & 9.05X & 22.71 & 1.17X            & 100\%        \\
(Yes, 0, 3)          & 6.73X & 16.89 & 1.17X            & 81\%         \\
(Yes, 0, 2)          & 4.64X & 11.64 & 1.17X            & 78.1\%       \\
(Yes, 0, 1)          & 2.29X & 5.75 & 1.22X            & 48.9\%       \\
\bottomrule
\end{tabular}
\caption{
\textbf{Cost and performance analysis for experiment setup two.}
}
\label{table:search_cost_res50}
\end{table*}

\begin{table*}[h!]
\centering 
\footnotesize
\begin{tabular}{r|rrrr}
\toprule
\multicolumn{1}{c|}{\begin{tabular}[c]{@{}r@{}}\textbf{Search Setting}\\ (Recurring, BSP runs, candidate runs)\end{tabular}} &
  \textbf{Search Cost} &
  \multicolumn{1}{c}{\begin{tabular}[c]{@{}r@{}}\textbf{Amortized}\\ (\# of recurrence)\end{tabular}} &
  \multicolumn{1}{c}{\begin{tabular}[c]{@{}r@{}}\textbf{Effective Training}\\ (vs. BSP) \end{tabular}} &
  \begin{tabular}[c]{@{}r@{}}\textbf{Success} \\ \textbf{Probability}\end{tabular} \\
\midrule 
Baseline: (No, 5, 5)      & 7.68X & 16.54 & 1.30X            & 100\%        \\
(No, 4, 4) & 6.14X & 13.22 & 1.30X            & 100\%        \\
(No, 3, 3)   & 4.61X & 9.93 & 1.30X            & 100\%        \\
(No, 2, 2)   & 3.07X & 6.61 & 1.30X            & 89.5\%       \\
(No, 1, 1)   & 1.54X & 3.32 & 1.30X            & 69.7\%       \\
(No, 1, 5)   & 3.67X & 7.90 & 1.63X            & 68.5\%       \\
(No, 1, 4)   & 3.14X & 6.76 & 1.59X            & 66.4\%       \\
(No, 1, 3)   & 2.61X & 5.62 & 1.53X            & 67.4\%       \\
(No, 1, 2)   & 2.07X & 4.46 & 1.49X            & 77.2\%       \\
(Yes, 0, 5)         & 2.68X & 5.77 & 1.87X            & 100\%        \\
(Yes, 0, 4)          & 2.14X & 4.61 & 1.87X            & 100\%        \\
(Yes, 0, 3)          & 1.67X & 3.60 & 1.87X            & 100\%        \\
(Yes, 0, 2)          & 1.07X & 2.30 & 1.87X            & 100\%        \\
(Yes, 0, 1)          & 0.54X & 1.16 & 1.87X            & 100\%        \\
\bottomrule
\end{tabular}
\caption{
\textbf{Cost and performance analysis for experiment setup three.}
}
\label{table:search_cost_cluster16}
\end{table*}

Table~\ref{table:search_cost_res50} details the search cost for the workload of ResNet50 and CIFAR-100 on an 8-worker cluster. Due to the relatively closer training throughput of BSP and ASP, the search cost is greater than the other, simpler, workload. Moreover, the search success probability is also lower for both recurring and new jobs. Reducing the search times to be four leads to a search cost of 9.05X of the BSP training and an amortized training cost of 22.71 recurring jobs.  The searching process also produces less effective training of 1.17X. In terms of a new training job, to ensure the success rate to be more than 99\%, running five times per setting is recommended, leading to a search cost of 17.86X amortized to 44.81 training sessions. However, even though the search cost is higher than the others for this setup, the effective training (1.12X) shows it is still more efficient than training with BSP.

Table~\ref{table:search_cost_cluster16} summarizes the cost and performance analysis for the workload of ResNet32 and CIFAR-10 with a cluster of size 16. We show that the search cost can be reduced to 0.54X the cost of BSP, cheaper than 1 BSP training session, due to only searched once. When facing a new training job, it is similarly safe to run three times per setting for guaranteed high success probability, for the cost of 4.61X of the BSP training. The search cost can be amortized with ten recurring jobs, similar to training the workload with a cluster of size 8. Additionally, \sysname can produce up to 1.87X of effective training compared to training with BSP.

\begin{figure*}[h]
    \centering
    \begin{subfigure}[t]{0.32\textwidth}
        \includegraphics[width=\textwidth]{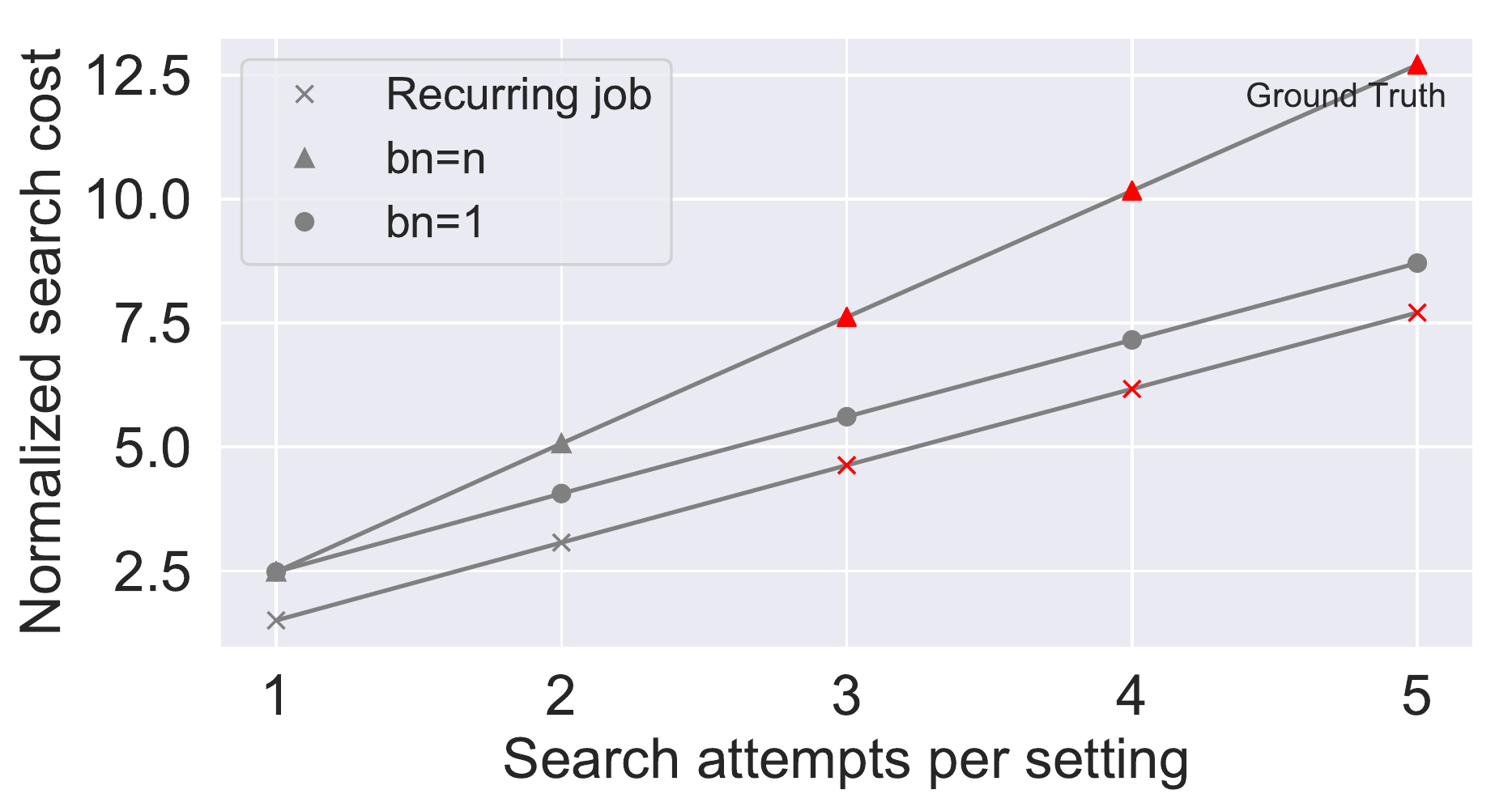}
        \caption{Exp. setup 1.}
        \label{fig:search_cost_eval_res32}
    \end{subfigure}
    \hfill
    \begin{subfigure}[t]{0.32\textwidth}
        \includegraphics[width=\textwidth]{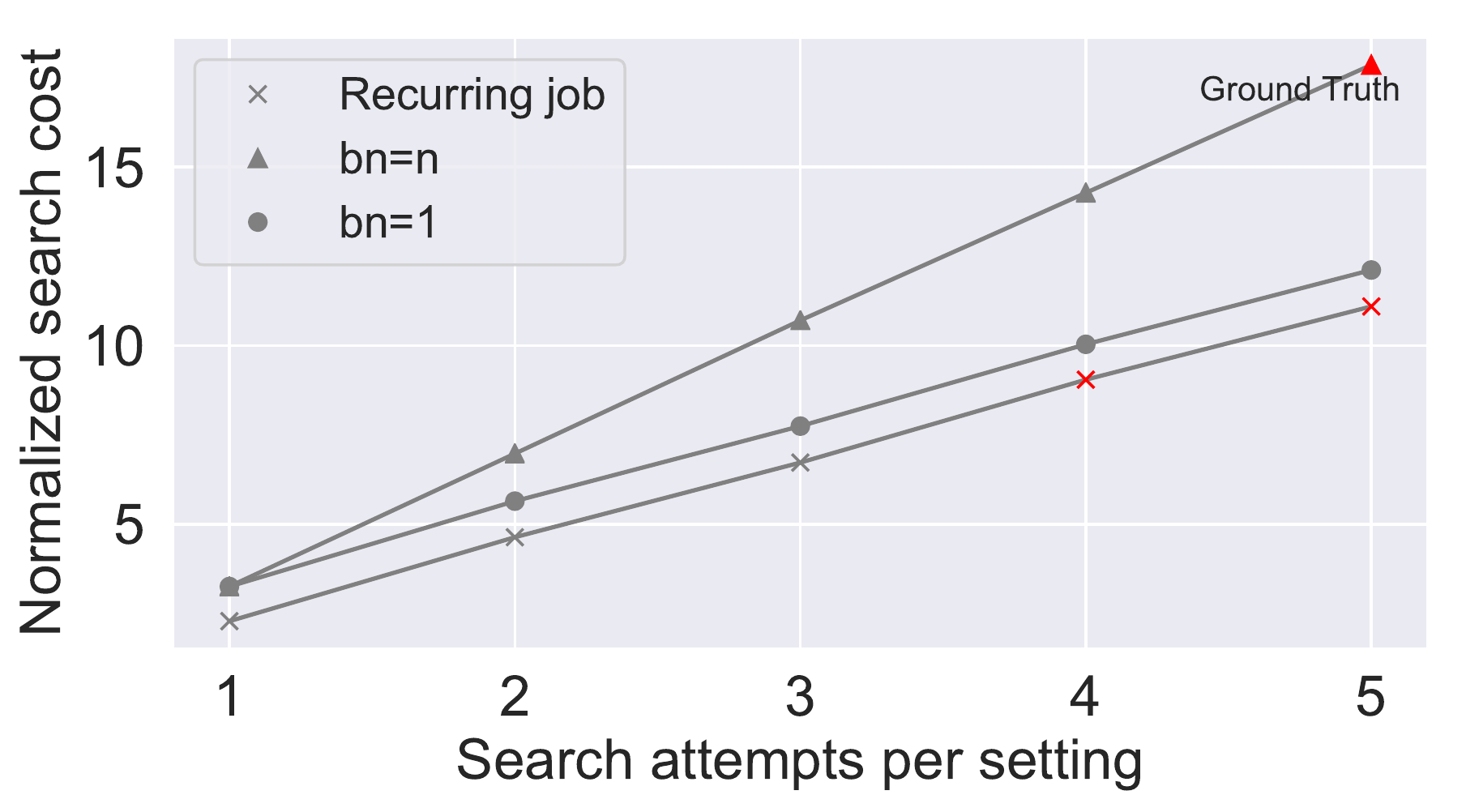}
        \caption{Exp. setup 2.}
        \label{fig:search_cost_eval_res50}
    \end{subfigure}
    \hfill
    \begin{subfigure}[t]{0.32\textwidth}
        \includegraphics[width=\textwidth]{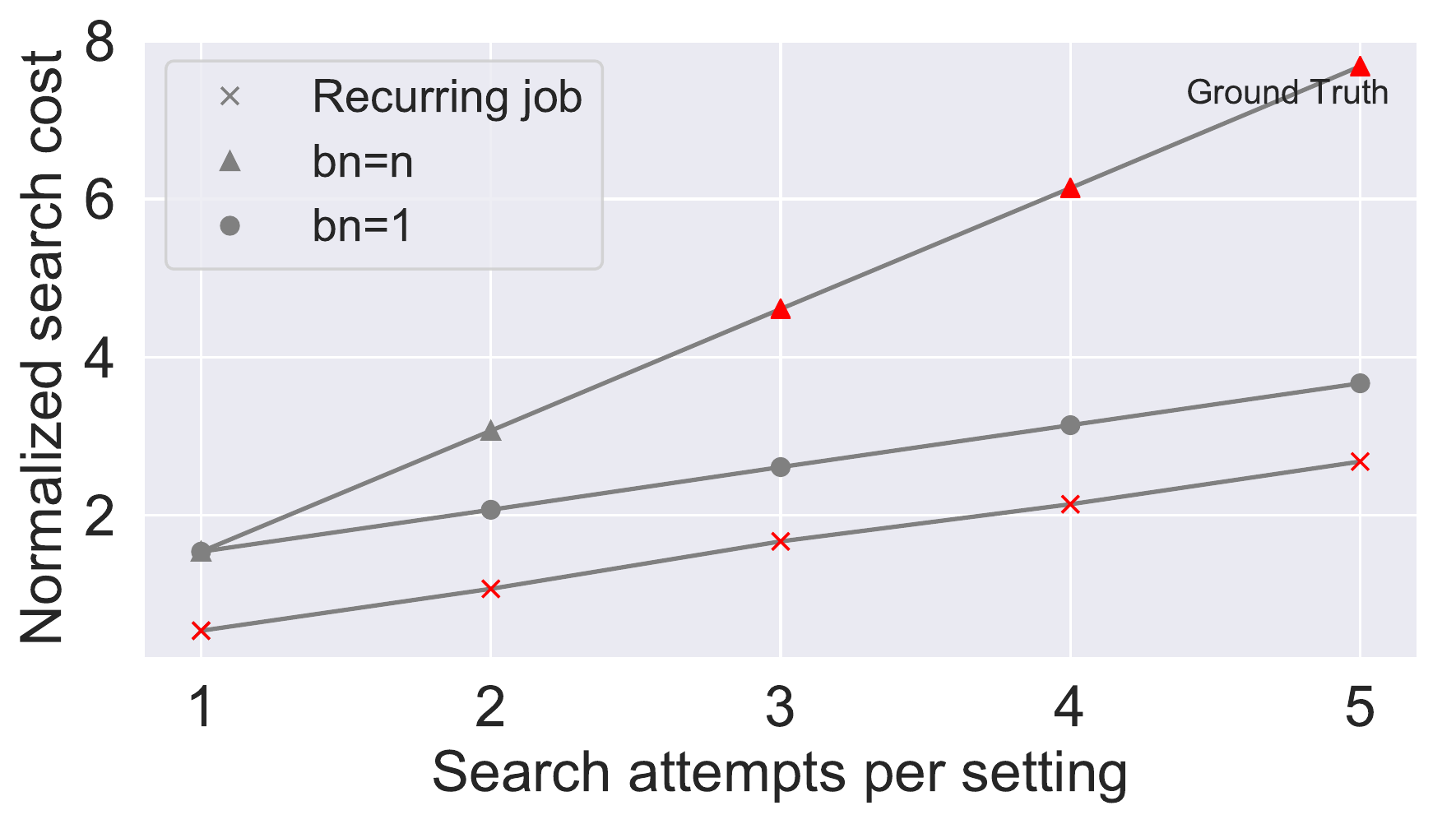}
        \caption{Exp. setup 3.}
        \label{fig:search_cost_eval_cluster16}
    \end{subfigure}
    \caption{\textbf{Search cost and performance trade-off.} We vary the number of measurement runs for each candidate switching timing for both recurring and new training jobs. Red markers denote the successful settings; we say a setting is successful if it obtains the ground-truth switch timing with $\geq 99\%$ probability.  
    }
    \label{fig:search_cost_eval}
\end{figure*}

Figure~\ref{fig:search_cost_eval} summarizes the normalized search cost to training with BSP for all three experiment setups.

\end{document}